\documentclass[11pt]{article}
\usepackage{amsmath,amsfonts,amsthm}
\newtheorem{theorem}{Theorem}[section]
\newtheorem{corollary}[theorem]{Corollary}
\newtheorem{lemma}[theorem]{Lemma}

\newtheorem{proposition}[theorem]{Proposition}
\newtheorem{definition}[theorem]{Definition}
\newtheorem{claim}[theorem]{Claim}

\def\defeq{\stackrel{\mathrm{def}}{=}}

\def\orig#1{\hat{#1}}
\def\ind#1{\left[#1 \right]}

\def\det#1{\textbf{det}\left(#1  \right)}

\def\typeuv{\mbox{Type}_{U,V} (A,\bb ,\cc )}

\def\prob#1#2{\Pr_{#1}\left[ #2 \right]}
\def\expec#1#2{\mbox{\bf E}_{#1}\left[ #2 \right]}

\def\norm#1{\left\| #1 \right\|}
\def\onenorm#1{\left\| #1 \right\|_{1}}
\def\infnorm#1{\left\| #1 \right\|_{\infty }}
\def\fnorm#1{\left\| #1 \right\|_{F}}
\def\setof#1{\left\{#1  \right\}}
\def\sizeof#1{\left|#1  \right|}

\def\dist#1#2{\mbox{{\bf dist}}\left(#1, #2 \right)}
\def\diff#1{\, d #1 \,}

\def\setminus{-}

\def\aa{\pmb{\mathit{a}}}
\newcommand\bb{\boldsymbol{\mathit{b}}}
\newcommand\cc{\boldsymbol{\mathit{c}}}

\renewcommand\gg{\boldsymbol{\mathit{g}}}

\newcommand\qq{\boldsymbol{\mathit{q}}}

\newcommand\xx{\boldsymbol{\mathit{x}}}
\newcommand\xxs{\boldsymbol{\mathit{x}}^{*}}

\newcommand\yy{\boldsymbol{\mathit{y}}}
\newcommand\yys{\boldsymbol{\mathit{y}}^{*}}

\newcommand{\xs}{x^{*}}
\newcommand{\ys}{y^{*}}

\def\ma#1{\mu_{A} (#1)}
\def\mb#1{\mu_{b} (#1)}
\def\mc#1{\mu_{c} (#1)}
\def\mbo#1{\mu_{b_{\Vb }} (#1)}
\def\mco#1{\mu_{c_{\Ub }} (#1)}
\def\mbi#1{\mu_{b_{V}} (#1)}
\def\mci#1{\mu_{c_{U}} (#1)}
\def\mai#1{\mu_{A_{V,U}} (#1)}

\def\origin{{\mbox{\boldmath $0$}}}

\def\vs#1#2#3{#1_{#2},\ldots , #1_{#3}}

\def\Vb{\bar{V}}
\def\Ub{\bar{U}}
\def\VUb{\overline{V,U}}

\def\Span#1{\mbox{{\bf Span}}\left(#1  \right)}
\def\abs#1{\left|#1  \right|}

\newdimen\pIR
\newcommand\StevesR{{\rm I\kern\pIR R}}
\def\Reals#1{\StevesR^{#1}}

\begin{document}

\title{Smoothed Analysis of Interior-Point Algorithms: Termination}

\author{
Daniel A. Spielman 
\thanks{Partially supported by NSF grant CCR-0112487.  
\texttt{spielman@math.mit.edu}}\\ 
Department of Mathematics\\ 
Massachusetts Institute of Technology
\and 
Shang-Hua Teng 
\thanks{
Partially supported by NSF grants CCR-9972532, and CCR-0112487. \texttt{steng@cs.bu.edu}}\\
Department of Computer Science\\
Boston University and\\
Akamai Technologies Inc.}

\maketitle

\begin{abstract}
We perform a smoothed analysis of the termination phase of
  an interior-point method.
By combining this analysis with the smoothed analysis of
  Renegar's interior-point algorithm in~\cite{DunaganSpielmanTeng},
  we show that the smoothed complexity of an interior-point algorithm
  for linear programming is
  $O (m^{3} \log (m/\sigma ))$.
In contrast, the best known bound on the worst-case complexity 
  of linear programming is
  $O (m^{3} L)$, where $L$ could be as large as $m$.
We include an introduction to smoothed analysis and a tutorial
  on proof techniques that have been useful in smoothed analyses.
\end{abstract}

\section{Introduction}

This paper has two objectives: to provide
  an introduction to smoothed analysis and to
  present a new result---the smoothed analysis
  of the termination of interior-point algorithms.
We begin with an intuitive introduction to smoothed
  analysis (Section~\ref{sec:informal}) followed
  by a more formal introduction (Section~\ref{sec:formal}).
After introducing necessary notation in Section~\ref{sec:notation},
  we survey the complexity of interior-point algorithms
  (Section~\ref{sec:ipa}), emphasizing the role of Renegar's condition
  number (Section~\ref{sec:condition}).
We then explain the termination algorithm (Section~\ref{sec:termination}),
  present its smoothed analysis at a high level (Section~\ref{sec:analDelta}),
  and then delve into the geometric (Section~\ref{sec:geometric})
  and probabilistic (Section~\ref{sec:probabilistic}) aspects
  of its analysis.
In Section~\ref{sec:probabilistic}, we include a tutorial of
  the fundamental techniques used in this work and
  in the smoothed analysis of the simplex method~\cite{SpielmanTengSimplex}.
Finally, in Section~\ref{sec:connection}, we explain how the
  analysis of  termination is related to the analysis of
  the simplex method.

\subsection{Intuitive Introduction to Smoothed Analysis}\label{sec:informal}
Folklore holds that most algorithms have much better
  performance in practice than can be proved theoretically.
This is partially due to the lack of a theoretical definition
  of ``practice'', partially due to the approximations made 
  in most theoretical analyses, and partially
  due to the dearth of performance measures considered in
  theoretical analyses.
In~\cite{SpielmanTengSimplex}, we suggested that smoothed
  analysis might provide a theoretically analyzable
  measure of an algorithm's performance
  that would be more predictive of its behavior in practice.
\footnote{We remark that a similar framework for 
  discrete problems was introduced by Blum and Spencer~\cite{BlumSpencer}}.  

Algorithms are typically analyzed through either
  worst-case or average-case complexity.
Worst-case analyses may disagree with practical experience
  because they are dominated by the most
  pathological input instances.
For many algorithms, these pathological inputs
  are rarely, if ever, encountered in practice,
  and are only known from lower-bound proofs.
In an attempt to create a less pessimistic analysis,
  researchers introduced
  average-case analysis, in which one defines a
  probability distribution on input instances and
  then measures the
  expected performance of an algorithm on
  inputs drawn from that distribution.
A low average-case complexity 
  provides some evidence that an algorithm
  may run quickly in practice.
However, this evidence is not conclusive
  as the inputs encountered by the algorithm
  in practice may not look like random inputs.

This discrepancy between theoretical 
  and experimental analysis 
  manifests itself in the analysis of
  linear programming algorithms.
The simplex method for linear programming is
  known to perform very well in practice, but
  to have exponential worst-case complexity
  \cite{KleeMinty,Murty,GoldfarbSit,Goldfarb,AvisChvatal,Jeroslow,AmentaZiegler}.
On the other hand, it is known to have polynomial
  average-case complexity under a number of
  notions of average-case
  \cite{Borg82,Borg77,SmaleRand,Haimovich,AdlerKarpShamir,AdlerMegiddo,ToddRand}.
Interior-point methods are known to have polynomial
  worst-case complexity~\cite{Karmarkar}.
However, their performance in practice is much
  better than their worst-case analyses would 
  suggest~\cite{LustigMarstenShanno3,LustigMarstenShanno2,AndersenIPM}.
It has been shown that the average-case complexity of
  interior-point methods is significantly lower
  than their worst-case complexity~\cite{AnstreicherJPY1,AnstreicherJPY2}
  (the term $L$ is replaced by $O (\log n)$);
  but these analyses are still a factor
  of approximately $\sqrt{n}$ off from that observed in practice.

Smoothed analysis provides an alternative
  to worst-case and average-case analyses,
  and also attempts to circumvent the
  need for a theoretical definition of 
  ``practical inputs''.
The smoothed complexity of an algorithm is
  defined to be the maximum over 
  its inputs of the expected running time of the
  algorithm under slight perturbations of that input.
The smoothed complexity is then measured
  as a function of the input size and the
  magnitude of the perturbation.
While many notions of perturbation are reasonable,
  most results have been obtained for Gaussian perturbations.
The assumption that inputs are subject to perturbation
  is reasonable in many circumstances:
  in many real-world numerical and geometric applications,
  data are
  derived from experimental and physical measurements and
  are therefore subject to errors
  \cite[paragraph 2, pp. 62]{Wilkinson}.
Perturbations can also be used to heuristically model the arbitrary decisions  
  that effect to formation of inputs that are presented to algorithms.

Two important aspects of smoothed analysis are:
\begin{itemize}
\item 
Smoothed analysis interpolates between worst-case
  and average-case analysis: 
By letting the magnitude of the
  random perturbation to the data 
  (\textit{e.g.}, the variance of the Gaussian noise)
  become large, one 
  obtains the traditional average-case complexity measure.
By letting the
  magnitude of the random perturbation go to zero, 
  one obtains the traditional
  worst-case complexity measure. 
In between, one obtains a model
  corresponding to noise in low-order digits of the input.

\item 
The smoothed complexity of an algorithm provides an upper
  bound on the expected complexity of the algorithm
  in every neighborhood of inputs.
That is, if the smoothed complexity of an algorithm is low,
  then it will run quickly on inputs drawn from any small
  neighborhood of inputs.

Thus, if the inputs presented to an algorithm in practice
  are subject to perturbation,
  the smoothed complexity of the algorithm should
  upper bound the practical performance of the algorithm.
\end{itemize}

In~\cite{SpielmanTengSimplex}, we introduced smoothed
  complexity by proving that a particular variant
  of the shadow-vertex simplex method has
  polynomial smoothed complexity.

\subsection{Formal Introduction to Smoothed Analysis}\label{sec:formal}

The inputs to most numerical and geometric problems 
  may be viewed as points in a vector space.
For example, an $m$ by $n$ real matrix 
  can be viewed as a vector in $\Reals{mn}$. 
Similarly, a set of $n$ points in $d$ dimensions 
  can be viewed as a vector in $\Reals{dn}$.

The most natural notion of perturbations of vectors
  in a real vector space is that of Gaussian perturbations.
Recall 
  that a Gaussian random variable with mean 0
  and variance $\sigma^{2}$ has density
\[
  \frac{1}{\sqrt{2 \pi} \sigma} e^{-x^{2}/ 2 \sigma^{2}},
\]
and that a 
  Gaussian random vector of variance $\sigma^2$ centered at
  the origin in $\Reals{n}$, denoted $\mathcal{N} (\origin, \sigma^{2})$,
  is a vector in which each
  entry is a Gaussian random variable of variance $\sigma^2$ and
  mean $0$, and has 
  density 
\[
  \frac{1}{\left(\sqrt{2 \pi} \sigma  \right)^{d} }
   e^{-\norm{\xx}^{2} / 2\sigma^{2}}.
\]

\begin{definition}[Gaussian perturbation]\label{def:gaussian}
Let $\orig{\xx} \in \Reals{n}$.
The Gaussian perturbation of 
  $\orig{\xx}$ of
variance $\sigma^{2}$
  is the random vector $\xx = \orig{\xx} + \gg$, where $\gg $ is
  a Gaussian random vector of variance $\sigma^{2}$, centered at the origin
  of $\Reals{n}$.
\end{definition}
The Gaussian perturbation of $\orig{\xx}$ may also be described
  as a Gaussian random vector of variance $\sigma^{2}$
  centered at $\orig{\xx}$.

Using the notion of Gaussian perturbation, we define
  the smoothed value of a function:
  
\begin{definition}[Smoothed value]\label{def:smoothed}
Let $f$ be a non-negative function on 
 $\Reals{n}$.
The smoothed value of $f$ with respect to Gaussian perturbations
  of variance $\sigma^{2}$ is
  given by
\[
 \max_{\orig{\xx}}
  \expec{\gg\leftarrow  \mathcal{N} (\origin, \sigma^{2}) }{f
   (\orig{\xx}  + \norm{\orig{\xx}} \gg )
  }
\]
\end{definition}

Note that in this definition we multiply the perturbation $\gg$
  by $\norm{\orig{\xx}}$ so that $\sigma$ represents
  the magnitude of the perturbation relative to the data.

\begin{definition}[Smoothed complexity]\label{def:smoothedComplex}
Let $A$ be an algorithm whose
  inputs can be expressed as vectors in $\Reals{n}$
and let $T_{A} (\xx)$ be the running time of algorithm $A$ on input 
  $\xx$.
Then
   the {\em smoothed complexity} of algorithm $A$ is
\[ \mathcal{C}_{A} (n, \sigma) = 
   \max_{\orig{\xx}\in \Reals{n}}\expec{\gg\leftarrow 
  \mathcal{N} (\origin , \sigma^{2}) }{T_{A} (\orig{\xx}+\norm{\orig{\xx}}\gg )}.
\]
\end{definition}

In \cite{SpielmanTengSimplex}, Spielman and Teng consider the
  complexity of a particular two-phase shadow-vertex simplex method
  on linear programs of 
  the form
\begin{eqnarray}
 &  \mbox{maximize} & \cc ^{T} \xx \nonumber \\
 & \mbox{subject to} & A \xx  \leq \bb,  \label{prg:A}
\end{eqnarray}
where $A$ is an $m$-by-$n$ matrix, $\bb $ is an $m$-vector,
  and $\cc $ is an $n$-vector.
They prove:

\begin{theorem}[Spielman-Teng]\label{the:st}
There is
  a two-phase shadow-vertex simplex method
  with time complexity $T (A, \bb ,\cc )$
  such that
 for every $m$-vector $\bb$ and $n$-vector $\cc $, 
  the smoothed complexity of the algorithm,
\[
 \max_{\orig{A}\in \Reals{m\times n}}
  \expec{G }{T \left(\orig{A}+\norm{\orig{A}}G, \bb ,\cc  \right)}
\]
is polynomial in $m$, $n$, and $1/\sigma $, 
  independent of $\bb$ and $\cc$,
  where $G$ is
  a Gaussian random $m$ by $n$ matrix of variance $\sigma^{2}$
  centered at the origin.
\end{theorem}

One need not limit smoothed analysis to measuring the
  expected complexity of algorithms in various neighborhoods. 
It is quite reasonable to prove other facts about
  the distribution of running times when the expectation
  does not exist, or when much stronger bounds can be proved.
For example, Blum and Dunagan~\cite{BlumDunagan} prove

\begin{theorem}[Blum-Dunagan]\label{thm:BD}
Let $\vs{\aa}{1}{n}$ be Gaussian random vectors in $\Reals{d}$ of
variance $\sigma^{2}< 1/ (2d)$ centered at points each of norm at most
$1$. 
Then, there exists a constant $c$ such that
  the probability that the perceptron 
  algorithm for linear programming
  takes
  more than  
$ \frac{c d^{3}n^{2}\log^{2 }
  (n/\delta)}{\delta^{2}\sigma^{2}}$  
iterations is at most $\delta$.
\end{theorem}

\section{Notation and Norms}\label{sec:notation}
Throughout the paper, we use bold letters such as
  $\bb$ and $\xx$ to denote vectors, capital letters
  such as $A$ and $G$ to denote matrices, and lower
  case letters to denote scalars.
In any context in which the vector $\bb$ is present,
  $b_{j}$ denotes the $j$th component of $\bb$.
For a set, $V$, we let $\bb_{V}$ denote the vector
  obtained by restricting $\bb$ to the coordinates
  in $V$.
When indexing and constructing matrices, we use
  the conventions of Matlab.
Thus, $A_{:,U}$ denotes the matrix formed by
  taking the columns indexed by $U$,
  and $A_{V,U}$ denotes the sub-matrix of 
  rows indexed by $V$ and columns
  indexed by $U$.
For sets, $U$ and $V$, we let
  $\Ub$ and $\Vb$ denote their complements.
We also let $\VUb$ denote the set of
  pairs $(i,j) \not \in (V,U)$;
 for example, we let $A_{\VUb }$ denote the set
  of entries of $A$ not in $A_{V,U}$.
For a matrix $A$ and a column vector $\bb$,
  we let $[A,\bb]$ denote the matrix obtained
  by appending column $\bb$ to $A$.

For an event, $\mathcal{E}$, we let
  $\ind{\mathcal{E}}$ denote the random variable that
  is 1 when $\mathcal{E}$ is true
  and is 0 otherwise.

We use of the following vector norms:
\begin{itemize}
\item $\norm{\xx} = \sqrt{\sum_{i} x_{i}^{2}}$,
\item $\onenorm{\xx} = \sum_{i} \abs{x_{i}}$, and
\item $\infnorm{\xx} = \max_{i} \abs{x_{i}}$,
\end{itemize}
and note that
\[
   \infnorm{\xx} \leq \norm{\xx} \leq \onenorm{\xx}.
\]
We also use the following matrix norms:
\begin{itemize}
\item $\norm{A} = \max_{\xx \not = 0} \norm{A \xx} / \norm{\xx}$, 
\item $\infnorm{A} = \max_{\xx \not = 0} \infnorm{A \xx} /
\infnorm{\xx}$, and
\item $\fnorm{A} = \sqrt{\mathbf{trace} (A^{T} A)}$, the square root
  of the sum of the squares of entries in $A$.
\end{itemize}
We note that
\begin{itemize}
\item $\infnorm{A} = \max_{i} \onenorm{A_{i,:}}$,
\item $\infnorm{A} \leq  \sqrt{n} \norm{A}$,

\item $\norm{A} \leq \fnorm{A}$, and

\item for sets $U$ and $V$, $\norm{A_{U,V}} \leq \norm{A}$.
\end{itemize}

\section{Complexity of Interior Point Algorithms}\label{sec:ipa}

A linear program is typically specified by a matrix $A$ together with
  two vectors $\bb $ and $\cc $, where 
  where $A$ is an $m$-by-$n$ matrix,
  $\cc $ is an $n$-dimensional row vector, and
  $\bb$ is an $m$-dimensional column vector.
There are several canonical forms of linear programs.
For the analyses in this paper,
  we will  consider linear programs of the form
\[
  \max \cc \xx \quad 
  \mbox{such that } A \xx \leq \bb , \quad \xx \geq \origin ,
\]
with dual
\[
  \min \yy \bb \quad 
  \mbox{such that } \yy A \geq \cc , \quad \yy \geq \origin.
\]
We will assume throughout that $m \geq n$.

If they exist,
  we denote the solutions to the primal and dual by
  $\xxs$ and $\yys$, and note that $\xxs$ is an $n$-dimensional column vector
  and $\yys$ is an $m$-dimensional row vector.

A linear programming algorithm should:
  (1) determine whether or not the linear program is feasible
  or bounded; 
  and, (2) if the program is feasible and bounded, output
  a solution.
One can either insist that the solution be a precisely
  optimal solution to the linear program, or merely
  a feasible point at which the objective function
  is approximately optimized.

The best bounds on the worst-case complexity of interior point
  methods, and for linear programming in general,
  were first obtained by Gonzaga~\cite{Gonzaga} 
  and Vaidya~\cite{Vaidya},
  who showed how to solve linear programs in $O (m^{3} L)$
  arithmetic operations%
\footnote{Vaidya's algorithm is somewhat faster as its
  complexity is $O ((m+n)n^{2} + (m+n)^{1.5}n)L$},
  where $m \geq n$ and
  $L$ is a parameter measuring the
  precision needed to perform the arithmetic operations exactly,
  and which here also appears in the number of arithmetic
  operations performed.
The definition of $L$ varies in the literature:
  Khachiyan~\cite{Khachiyan}, Karmarkar~\cite{Karmarkar},
  and Vaidya~\cite{Vaidya} define $L$ 
  for integer matrices $A$ to
  be some constant times
\begin{multline*}
  \log (\mbox{largest absolute value of the determinant
   of any square sub-matrix of $A$})\\
 + 
  \log (\infnorm{\cc})
 +
  \log (\infnorm{\bb})
 +
  \log (m + n).
\end{multline*}
Under  this definition, $L$ is not efficiently
  computable, and unless $A$ 
  comes from a very special class of matrices,
  it is difficult to find $L$ below
  $\Omega (n)$.
Others use cruder bounds such as
  the total number of bits in a row of the matrix
  or the total number of bits in the entire 
  matrix~\cite{Wright}.

To understand the time complexity of interior point
  algorithms, we note that they
  are typically divided into three phases:
\begin{itemize}
\item [] [Initialization]: In this phase,
  the algorithm determines whether or not the program is
  feasible and bounded; and, if it is feasible and
  bounded, returns a feasible point.
\item [] [Iteration]: In this phase, the algorithm
  iteratively finds feasible points 
  on which the objective function becomes
  increasingly closer to optimal.

\item [] [Termination]: In this phase, the algorithm
  jumps from a feasible point that is close to optimal
  to the exact optimal solution of the linear program.
\end{itemize}
Of course, if one merely desires an approximate solution 
  to the linear program, then one can skip the termination phase.
However, the dependency on $L$ appears in both the initialization
  and termination phase. 
So, the worst-case complexity of linear programming algorithms
  is not decreased by merely asking for an approximate solution.

The kernel of an interior-point algorithm is the iteration phase,
  in which feasible points of increasing quality are computed.
A typical measure of quality in a primal algorithm 
  is the \textit{optimality gap} between
  the objective function at the current point and the optimal, while
  in a primal-dual algorithm it is the \textit{duality gap} between
  the current primal and dual feasible points.
In either case, one can prove that
  after $k$ iterations the
  gap decreases by the multiplicative
  factor $\left(1 - \frac{c}{\sqrt{m}} \right)^{k}$,
  for some constant 
  $c$~\cite{RenegarAlg,Vaidya,YeBook}.
If performed carefully, each
   of these iterations has complexity
   $O (m^{5/2})$~\cite{Gonzaga}.
Therefore, the total number of arithmetic operations 
  required to reduce
  the gap from $R$ to $\epsilon $ is
  $O (m^{3}\log (R/\epsilon))$.
The worst-case complexity bounds come from the facts
  that 
  a typical interior-point algorithm discovers
  a feasible point with initial gap
  bounded by $R = 2^{O (L)}$ in the initialization phase, 
  and requires
  a point with gap less than
  $\epsilon = 2^{-O (L)}$ to start the termination phase.

In practice, the speed of interior point methods
  is much better than that proved in their worst-case 
  analyses~\cite{LustigMarstenShanno3,LustigMarstenShanno2,AndersenIPM}.
This difference in speed seems to have two sources:
  first, the upper bound of $L$ is overly pessimistic;
  and, second, 
  the improvement made at each iteration is typically
  much better than $\left(1-\frac{c}{\sqrt{m}} \right)$.
However, we note that
  Todd~\cite{ToddLower} and Todd and Ye~\cite{ToddYeLower}
  have exhibited linear programs in which
  $\Omega (n^{1/3})$ iterations are required to improve
  the gap by a constant factor.

In this paper, we perform a smoothed analysis of
  a simple termination phase for 
  interior point methods.
By combing this analysis with the analysis
  of the first two phases of 
  Renegar's interior point algorithm~\cite{RenegarFunc}
  in~\cite{DunaganSpielmanTeng},
  we obtain an interior point algorithm with
  smoothed complexity
  $O (m^{3} \log (m/\sigma))$.
Essentially, this analysis replaces the dependency
  on $L$ in the initialization and termination phases
  with a dependency on $\log (m/\sigma)$. 
We conjecture that one can improve 
  this smoothed complexity estimate by
  proving that the smoothed
  number of iterations taken by an interior
  point method is less than
  $O (\sqrt{m} \log (m / \sigma))$.

Renegar~\cite{RenegarFunc,RenegarCond,RenegarPert}
  defined a condition number $C (A,\bb ,\cc)$
  of a linear program, and developed an
  algorithm for the initialization
  of an interior point method that
  runs in time $O (m^{3} \log (C (A,\bb ,\cc )))$
  and returns a feasible point with
  initial optimality gap $R \leq O (m  C (A, \bb ,\cc ))$.
Applying a primal iteration phase to this feasible point,
  one obtains an algorithm that
  after $O (\sqrt{m} \log (m C (A, \bb ,\cc )) / \epsilon )$
  rounds and
  $O (m^{3} \log (m C (A, \bb ,\cc )) / \epsilon )$
  arithmetic operations
  produces points with optimality gap at most $\epsilon$.
Renegar's condition number will be discussed further
  in Section~\ref{sec:condition}.

Dunagan, Spielman and Teng~\cite{DunaganSpielmanTeng} perform a smoothed
  analysis of Renegar's condition number and prove:

\begin{theorem}[Dunagan-Spielman-Teng]\label{thm:DST}
Let $\orig{A}$ be an $m$-by-$n$ matrix for $m \geq n$,
  $\orig{\bb} $ an $m$-vector,
  and  $\orig{\cc}$ an $n$-vector for which
 $\fnorm{\orig{A}, \orig{\bb }, \orig{\cc }} \leq 1$,
and let $A$, $\bb$ and $\cc$
  be the Gaussian perturbations of $\orig{A}$,
  $\orig{\bb}$ and $\orig{\cc}$ of variance 
  $\sigma \leq 1/\sqrt{mn}$.
Then, 
\[
\expec{A, \bb ,\cc}{\log (C (A, \bb ,\cc))}
\leq 
O (\log (m / \sigma)).
\]
\end{theorem}
Combining this analysis with that of Renegar,
  we find that the smoothed complexity of finding
  an $\epsilon$-optimal solution to
  a linear program is
  $O (m^{3} \log (m / \sigma \epsilon ))$.

In Section~\ref{sec:termination}, we define a
  simple termination algorithm that takes
  $O (m^{3})$ arithmetic operations.
We define $\delta (A, \bb , \cc)$ to be the
  greatest number such that
  $\cc \xxs - \cc \xx \leq \delta (A, \bb ,\cc )$
  implies that the termination algorithm is successful.
Thus, after $O (\sqrt{m} \log (m C (A, \bb ,\cc))/ \delta (A, \bb ,\cc ))$
  iterations,
  and  $O (m^{3} \log (m C (A, \bb ,\cc))/ \delta (A, \bb ,\cc ))$
  arithmetic operations, one can apply the
  termination phase to find the exact solution to
  the linear program.
Like Karmarkar~\cite{Karmarkar}, we
  handle the technical difficulty that the algorithm does
  not know $C (A, \bb , \cc)$ or
  $\delta (A, \bb , \cc )$
  by periodically attempting to terminate,
  but only once every 
  $\sqrt{n}$ iterations so 
  as not to increase the complexity of the algorithm.

In Theorem~\ref{thm:delta},
  proved over Sections~\ref{sec:analDelta}, \ref{sec:geometric} and
  \ref{sec:probabilistic}, we show that
  the smoothed value of\linebreak
  $\max \left(1,\log \left(1/\delta (A, \bb  ,\cc ) \right) \right)$
  is $O (\log (m / \sigma ))$.
We thus prove:

\begin{theorem}[Smoothed Complexity of IPM]\label{thm:MainPaper}
Let $\orig{A}$ be an $m$-by-$n$ matrix for $m \geq n$ ,  $\orig{\bb} $ an $m$-vector,
  and  $\orig{\cc}$ an $n$-vector for which
 $\fnorm{\orig{A}, \orig{\bb }, \orig{\cc }} \leq 1$,
and let $A$, $\bb$ and $\cc$
  be the Gaussian perturbations
  of $\orig{A}$, $\orig{\bb}$ and $\orig{\cc}$
  of variance $\sigma < 1/\sqrt{mn}$.
Let $T (A,\bb,\cc)$ denote the complexity 
  of Renegar's interior point algorithm 
  with the periodic application of the termination
  procedure described in Section~\ref{sec:termination}.
Then,
\[
\expec{A, \bb ,\cc}{T (A,\bb ,\cc )}
\leq O (m^{3}\log (m / \sigma)).
\]
\end{theorem}

While this is the statement of the complexity that is most
  natural for our proof techniques, we note that it is
  not exactly the form specified in
  Definition~\ref{def:smoothedComplex}.
The difference comes from the upper bounds on $\sigma$
  and  $\fnorm{\orig{A}, \orig{\bb }, \orig{\cc }}$ in the
  statement of the theorem.
As the behavior of the interior point methods are unchanged by
  multiplicative changes to $\orig{A}$, $\orig{\bb}$
  and $\orig{\cc}$,
  only the upper bound on $\sigma$ is significant:
  if $\fnorm{\orig{A}, \orig{\bb }, \orig{\cc }} \geq 1$, then one
  can scale down $\orig{A}$, $\orig{\bb}$, $\orig{\cc}$, and $\sigma$
  to make $\fnorm{\orig{A}, \orig{\bb }, \orig{\cc }} = 1$.
One could adjust Theorem~\ref{thm:MainPaper}
  in two ways to handle $\sigma > 1/\sqrt{mn}$:
  one could either extend the proofs, or one
  could use Theorem~\ref{thm:MainPaper} as a black-box 
  and derive the more general statement from it.
Such a proof could proceed by observing that
  a Gaussian of variance $\sigma^{2}$ is the sum of
  a Gaussian of variance $\tau^{2}$ and a Gaussian of
  variance $\sigma^{2} - \tau^{2}$.
Thus, one can apply Theorem~\ref{thm:MainPaper} with a
  Gaussian of variance $\tau^{2}$ to the result
  of perturbing the original data by a Gaussian
  of variance $\sigma^{2} - \tau^{2}$, for an appropriate
  choice of $\tau$.
The reader can find a precise implementation of this
  technique in ~\cite[Section 5.1]{SpielmanTengSimplex}.

\section{Renegar's Condition Number for Linear Programming}\label{sec:condition}

In an effort to develop a parameter in which to measure
  the complexity of linear programming that was more
  natural than $L$,
Renegar~\cite{RenegarFunc,RenegarCond,RenegarPert}, 
  introduced the condition
  number, $C (A,\bb,\cc)$, of a linear program
  and developed an interior point method
  that runs in time
  $O (m^{3} \log (C (A, \bb ,\cc) / \epsilon ))$.
In contrast with the parameter $L$, $C (A,\bb ,\cc)$
  is naturally defined for rational or
  real matrices $A$.
Moreover, $C (A, \bb ,\cc)$ is often much smaller
  than $L$.

Formally,
  we define the distance of a linear program
  specified by $(A, \bb ,\cc)$ to primal ill-posedness
  to be
\[
\kappa_{P} (A, \bb )
 = 
   \left\{
\begin{array}{l}
  \sup \setof{\kappa : 
  \fnorm{A-A',\bb -\bb '} \leq \kappa 
  \text{ implies }
  A' \xx \leq \bb , \xx \geq 0
  \text{ is feasible}}\\
 \qquad  \text{if   $A' \xx \leq \bb , \xx \geq 0$ is feasible, and}\\
    \sup \setof{\kappa : 
  \fnorm{A-A',\bb -\bb '} \leq \kappa 
  \text{ implies }
  A' \xx \leq \bb , \xx \geq 0
  \text{ is infeasible}}\\
 \qquad \text{if   $A' \xx \leq \bb , \xx \geq 0$ is infeasible.}\\
\end{array}
 \right. 
\]
The distance to dual ill-posedness, $\kappa_{D} (A, \cc)$,
  is defined similarly.
We then define
  $C (A,\bb,\cc)$
  to be the maximum of
  the primal condition number $C_{P} (A,\bb)$
  and the dual condition number $C_{D} (A,\cc )$,
  where
  $C_{P} (A, \bb )$ and $C_{D} (A, \cc )$ are the normalized
  distances to primal and dual ill-posedness:
\[
  C_{P} (A, \bb) = 
  \fnorm{A,\bb } / \kappa_{P} (A, \bb )
\text{ and }
  C_{D} (A, \cc) = 
  \fnorm{A,\cc } / \kappa_{D} (A, \cc ).
\]
We remark that, with this normalization,
  $C_{P}$ and $C_{D}$ are always at least 1.

We also note that the linear programs for which
  Todd~\cite{ToddLower} and Todd and Ye~\cite{ToddYeLower}
  prove a   $\Omega (n^{1/3})$ iteration lower bound
  have exponentially poor condition.
It is not known if one can prove such an iteration lower
  bound for a well-conditioned linear program.
 
\section{Termination}\label{sec:termination}
One can often terminate linear programming algorithms
  that approach the optimal solution of a linear program
  by using a good solution to guess the optimal solution.
The process by which this is done is often called
  termination or rounding.
Termination is possible because at the optimal solution
  a number of the inequalities are tight, and the knowledge
  of the identity of these inequalities is enough to reconstruct
  the optimal solution.
Thus, most termination algorithms work by guessing
  that the inequalities having the least slack at a very good
  solution are those which have no slack at the optimal solution.

We being by recalling the facts that we will use to
  prove that termination is possible, ignoring complications
  that may occur with probability zero for perturbed
  $A$, $\bb$ and $\cc$.
We begin with
\begin{proposition}\label{pro:prob1}
For Gaussian distributed $A$, $\bb$ and $\cc$,
  with probability 1, the program specified by
  $(A, \bb ,\cc)$ is either infeasible, unbounded,
  or has unique primal and dual optimal solutions,
  $\xxs$ and $\yys$.
Moreover, $\xxs$ makes tight exactly
  $n$ of the inequalities
  $\setof{x_{i} \geq 0} \cup \setof{A_{j,:} \xx \leq b_{j}}$
  and  $\yys$ makes tight exactly
  $m$ of the inequalities
  $\setof{y_{j} \geq 0} \cup \setof{\yy A_{:,i} \geq c_{i}}$.
\end{proposition}
\begin{proof}
If the primal program is feasible and bounded but does not
  have a unique optimal solution, then the space of optimal
  solutions must lie in a subspace defined by fewer than
  $n$ of the inequalities
  $\setof{x_{i} = 0} \cup \setof{A_{j,:} \xx = b_{j}}$,
  and $\cc$ must be orthogonal to this subspace.
However, as this restricts $\cc$ to a set of measure zero
  and the number of such possible subspaces is finite
  given $A$ and $\bb$, this is an event with probability zero.
By symmetry, the same holds for the optimal solution
  of the dual program.
To prove the second part, we note that if
  $n+1$ of the inequalities are tight at $\xxs$,
  then these inequalities form a system
  of $n+1$ equations in $n$ variables
  that has a solution. 
As any such degeneracy has probability zero, and
  there are only finitely many such possible degeneracies,
  the probability of this happening is zero.
\end{proof}

We now recall the Duality Theorem of Linear Programming:

\begin{theorem}[LP duality]\label{thm:lpDuality}
For a linear program specified by
  $(A, \bb ,\cc )$,
\begin{itemize}
\item (Weak Duality) for every primal feasible $\xx$
  and dual feasible $\yy$, $\yy \bb \geq \cc \xx  $, and
\item (Strong Duality) if the linear program is bounded
  and feasible then
  for primal optimal $\xxs$ and a dual optimal $\yys$, we have
  $\yys \bb = \yys A \xxs = \cc \xxs  $.
\end{itemize}
\end{theorem}

For a feasible and bounded linear program $(A ,\bb ,\cc)$ with
  unique optimal primal and dual solutions $\xxs$ and $\yys $, 
  we define
\begin{align*}
U & = \{i:\xx_{i}^{*} > 0 \}\\
V & = \{j:\yy_{j}^{*} > 0 \},
\end{align*}
and we say that the program is of type $(U,V)$.

We can show that $U$ and $V$ are related to the
  set of tight constraints:

\begin{lemma}[Tight constraints]\label{lem:slackness}
For a feasible and bounded linear program specified by 
 $(A, \bb , \cc)$, we have
\begin{align*}
V & \subseteq  \{j: A_{j,:}\xxs = \bb_{j} \}, \text{ and }\\
U & \subseteq  \{i: \yys A_{:,i} =\cc_{i} \}.
\end{align*}
\end{lemma}
\begin{proof}
Let 
\begin{align*}
V' & = \{j: A_{j,:}\xxs = \bb_{j} \}, \text{ and }\\
U' & = \{i: \yys A_{:,i} =\cc_{i} \}.
\end{align*}
To show that $U\subseteq U'$, assume
  by way of contradiction that there exists an $i\in U$
  such that $\yys A_{:,i} >\cc_{i}$.
Because $\xs_{i} >0 $ we have 
  $\yys A_{:,i}\xs_{i}  >\cc_{i}\xs_{i}$, which would imply
  $\yys \bb > \cc \xxs  $ and contradict Theorem~\ref{thm:lpDuality}.
Therefore $U \subseteq U'$. 
We can similarly show that $V \subseteq V'$.
\end{proof}

With probability 1, these sets are actually identical:

\begin{lemma}[$U$ and $V$]\label{lem:samesize}
For Gaussian distributed $A$, $\bb$ and $\cc$,
  if the corresponding linear program is bounded and
  feasible, then with probability 1, 
\begin{align*}
V & = \{j: A_{j,:}\xxs = \bb_{j} \}, \text{ and }\\
U & =  \{i: \yys A_{:,i} =\cc_{i} \}.
\end{align*}
\end{lemma}
\begin{proof}
Define $V'$ and $U'$ as in the proof of Lemma~\ref{lem:slackness}.
We will show
 $\sizeof{U}=\sizeof{U'}$ and $\sizeof{V}= \sizeof{V'}$.
By Proposition~\ref{pro:prob1}, 
  with probability 1,
  the number of zeros in $\xxs$
  plus
  $\sizeof{U'}$ equals $n$. 
Because $\xxs $ is an $n$-place vector, the number of zeros in
  $\xxs $ plus the number of non-zeros in $\xxs $, which is $\sizeof{U}$, is
  equal to $n$.
Thus $\sizeof{U}=\sizeof{U'}$.
Similarly, $\sizeof{V}=\sizeof{V'}$.
\end{proof}

We will consider the following termination scheme: suppose $\xx $ is an
  approximate solution to the primal program, 
  we let $U(\xx )$ and $V (\xx)$ be the set of indices such that
\[
\{\xx_{i} : i \not \in U(\xx )\} \cup \{\bb_{j} - A_{j,:}\xx: j\in V (\xx )\}
\]
are the smallest $n$ values in 
$\{\xx_{i}\} \cup \{\bb_{j} - A_{j,:}\xx\}$.
We then guess the optimal solution to be
   the solution to the following linear
  system: $\xx_{i} = 0$ for $i \not \in U(\xx )$ and 
  $\bb_{j} - A_{j,:}\xx = 0$ for $j\in V (\xx )$.
We will show that if $\xx$ is sufficiently close
  to optimal, then this termination scheme produces
  the optimal solution to the linear program.
We now define $\delta (A ,\bb ,\cc)$ to measure
  how close to optimal $\xx$ needs to be.

\begin{definition}[$\delta (A, \bb ,\cc )$]
For a feasible and bounded linear program specified by $A$, $\bb$ and $\cc$,
  we define
\[
\delta (A, \bb , \cc)
\]
to be the supremum of the $\delta $ for which
\begin{equation}\label{}
   (\cc \xxs - \cc \xx) < \delta 
 \text{ implies }
  U (\xx) = U 
 \text{ and }
  V (\xx) = V .
\end{equation}
For unbounded or infeasible programs, we set $\delta$ to $\infty$.
\end{definition}

The main technical contribution of this paper is:
\begin{theorem}\label{thm:delta}
Let $\orig{A}$ be an $m$-by-$n$ matrix ,  $m \geq n$, $\orig{\bb} $ an $m$-vector,
  and  $\orig{\cc}$ an $n$-vector for which
 $\norm{\orig{A}},  \norm{\orig{\bb }}, \norm{\orig{\cc }} \leq 1$,
and let $A$, $\bb$ and $\cc$
  be a Gaussian random matrix and two Gaussian random vectors
  of variance $\sigma^{2}$ centered at $\orig{A}$, $\orig{\bb}$
  and $\orig{\cc}$, respectively.
Then, for $\sigma^{2} \leq 1$,
\[
  \expec{}{
\max \left(1, 
\log \left(\frac{1}{\delta (A, \bb ,\cc )} \right)  \right)}
\leq 
 O ( \log (m  /\sigma )).
\]
\end{theorem}

Our proof of Theorem~\ref{thm:delta} is broken
  into three sections.
In Section~\ref{sec:analDelta} we
  define geometric quantities that we will use
  to bound $\delta (A, \bb ,\cc)$,
  state the relation between $\delta (A , \bb ,\cc)$
  and these quantities proved in
  Section~\ref{sec:geometric},
  and state the probability bounds
  for these quantities obtained in
  Section~\ref{sec:probabilistic}.
The rest of the material in Section~\ref{sec:analDelta}
  is a routine calculation using the
  results of Sections~\ref{sec:geometric},
  \ref{sec:probabilistic} and
  Theorem~\ref{thm:DST}.
The reader will probably be most interested
  in Section~\ref{sec:probabilistic},
  which we begin with an intuitive explanation
  of how the probability estimates are obtained,
  carefully explain the tools used to make these
  arguments rigorous, and then finally apply
  these tools to obtain the probability bounds.
  
We remark that Theorem~\ref{thm:delta} depends very little
  on the properties of Gaussian random variables.
Aside for the bound on $\expec{}{\log{\norm{A}}}$
  of Proposition~\ref{pro:expecNorm}, which is easily generalized
  to other distributions, the only fact about Gaussian random variables
  used is that proved in Lemma~\ref{lem:GaussianSmooth}.
Thus, one could prove statements similar to Theorem~\ref{thm:delta}
  for a number of families of perturbations.

\section{Smoothed Analysis of $\log (1/\delta) $}\label{sec:analDelta}
Our analysis of the probability that $\delta (A,\bb ,\cc)$
  is small will be divided into two parts:
  a geometric condition for $\delta (A, \bb ,\cc)$ to be small,
  and a bound on the probability
  that this geometric condition is satisfied.

To described the geometric condition, we define
  the following five quantities for
  bounded and feasible linear programs
  with unique optimal primal and dual solutions
  $\xxs$ and $\yys$.
\begin{itemize}
\item $\alpha_{P} (A, \bb, \cc) = \min_{i \in U} \xs_{i} $,
\item $\alpha_{D} (A, \bb, \cc) = \min_{j \in V} \ys_{j} $,
\item $\beta_{P} (A, \bb, \cc) = \min_{j \in \Vb } b_{j} - A_{j,:} \xxs$,
\item $\beta_{D} (A, \bb, \cc) = \min_{j \in \Ub } \yys A_{:,i} - c_{i}$,
\item $\gamma (A ,\bb , \cc ) = \min_{k \in U}
  \dist{A_{V,k}}{\Span{A_{V , U \setminus k}}}$.
\end{itemize}
The geometric condition is that one of these
  five quantities is small.

When $A$, $\bb$ and $\cc$
  are clear from context, we will just write
  $\alpha_{P}$, $\alpha_{D}$, $\beta_{D}$, $\beta_{P}$, $\gamma$
  or $\delta$.
Note that 
\[
  \alpha_{D} (A ,\bb, \cc  ) = 
  \alpha_{P} (-A^{T}, -\cc^{T}, -\bb^{T})
\quad \text{ and } \quad 
  \beta_{D} (A ,\bb, \cc  ) = 
  \beta_{P} (-A^{T}, -\cc^{T}, -\bb^{T}).
\]

In Section~\ref{sec:geometric}, we prove

\begin{lemma}\label{lem:canRound}
For a linear program specified by $(A, \bb ,\cc)$ 
  with unique optimal primal and dual solutions $\xxs$ and $\yys$,
  let 
\[
\lambda (A,\bb ,\cc  ) = \min \left(\alpha_{P} (A,\bb ,\cc ),
	\alpha_{D} (A,\bb ,\cc ),
	\beta_{P} (A,\bb ,\cc ),
	\beta_{D} (A,\bb ,\cc ) \right).
\]
Then,
\[
  \delta (A, \bb ,\cc )
\geq 
  \frac{
   \lambda (A, \bb , \cc )^{2} \gamma (A, \bb , \cc )
   }{
   2 \max (1,\sqrt{n} \norm{A}) \left(1 + \norm{A} \right)
   }.
\]
\end{lemma}

We define $\mathcal{F} (A, \bb ,\cc)$ to be the event that
  the linear program specified by $A, \bb , \cc$
  is feasible and bounded.
In Section~\ref{sec:probabilistic}, we prove

\begin{lemma}[Probability of small $\alpha $]\label{lem:probAlpha}
Under the conditions of Theorem~\ref{thm:delta},
\[
  \prob{}
    {\alpha_{P} (A, \bb ,\cc) \leq 
            \frac{\epsilon}{\left(\norm{A}+2 \right)^{2} 
                      \left(\norm{\xxs}+ 1 \right)}
   \text{ and } \mathcal{F} (A, \bb , \cc )}
\leq 
  \frac{8 \epsilon n (m+1)}
       {\sigma^{2}}.
\]
\end{lemma}

\begin{lemma}[Probability of small $\beta $]\label{lem:probBeta}
Under the conditions of Theorem~\ref{thm:delta},
\[
  \prob{}{\beta_{P} (A, \bb , \cc) 
   \leq  \frac{\epsilon }{\max \left(1,\norm{A} \norm{\xxs} \right)}
     \text{ and } \mathcal{F} (A, \bb , \cc) )
}
\leq 
  \frac{4 \epsilon m }{\sigma^{2}}.
\]
\end{lemma}

\begin{lemma}[Probability of small $\gamma $]\label{lem:probGamma}
Under the conditions of Theorem~\ref{thm:delta},
\[
  \prob{A_{V,U}}
       {\gamma (A, \bb ,\cc ) \leq 
   \frac{\epsilon }{\left(1 + \norm{\xxs}^{2} + \norm{\yys}^{2} \right)
                    \left(\norm{A} + 3 \right)}
   \text{ and } \mathcal{F} (A,\bb ,\cc )
 } \leq 
\frac{
  \epsilon n e 
}{
  \sigma^{2}
}.
\]
\end{lemma}

From these three lemmas, we can reduce our analysis of
  the probability that $\delta (A, \bb, \cc)$ is small
  to an analysis of the probability that
  $\norm{\xxs}$, $\norm{\yys}$ or $\norm{A}$
  is large.

\begin{lemma}\label{lem:reduce}
Under the conditions of Theorem~\ref{thm:delta},
\begin{equation}\label{eqn:reduce}
\prob{A, \bb ,\cc }
 {\delta (A, \bb ,\cc) 
 \left(\norm{A} + 3 \right)^{7}
  \left(1 + \norm{\xxs} + \norm{\yys } \right)^{4}
\leq 
\epsilon }
\leq 
  \frac{21 \epsilon^{1/3} n^{1/6} (n + 1) (m+1)}
       {\sigma^{2}}.
\end{equation}
\end{lemma}
\begin{proof}
As $\delta (A, \bb ,\cc)$ is infinite for infeasible or unbounded
  programs, 
\begin{multline*}
\prob{A, \bb ,\cc }
 {\delta (A, \bb ,\cc) 
 \left(\norm{A} + 3 \right)^{7}
  \left(1 + \norm{\xxs} + \norm{\yys } \right)^{4}
\leq 
\epsilon }\\
=
\prob{A, \bb ,\cc }
 {\delta (A, \bb ,\cc) 
 \left(\norm{A} + 3 \right)^{7}
  \left(1 + \norm{\xxs} + \norm{\yys } \right)^{4}
\leq 
\epsilon \text{ and } \mathcal{F} (A, \bb ,\cc )}.
\end{multline*}

As $\alpha_{D} (A, \bb , \cc) = \alpha_{P} (-A^{t}, -\cc^{t},-\bb^{t})$,
 and
\[
            \frac{\epsilon}{\left(\norm{A}+2 \right)^{2} 
                      \left(\norm{\xxs}+ 1 \right)}
\leq 
 \frac{\epsilon }{\max \left(1,\norm{A} \norm{\xxs} \right)},
\]
  Lemmas~\ref{lem:probAlpha} and~\ref{lem:probBeta} imply
\begin{align*}
&  \prob{A,\bb ,\cc}
    {\min \left(\alpha_{P},\alpha_{D},\beta_{P},\beta_{D} \right)
     \leq 
            \frac{\epsilon}{\left(\norm{A}+2 \right)^{2} 
                 \left(\max \left(\norm{\xxs}, \norm{\yys} \right)
                   + 1 \right)}
 \text{ and } \mathcal{F} (A, \bb ,\cc ) }\\
& \leq 
  \frac{8 \epsilon n (m+1) + 8 \epsilon m (n+1) +4 \epsilon m + 4 \epsilon n}
       {\sigma^{2}}\\
& \leq 
  \frac{8 \epsilon \left( (n + 1/2) (m+1) + (m + 1/2) (n+1) \right)}
       {\sigma^{2}}.
\end{align*}
Let $\lambda = \min \left(\alpha_{P},\alpha_{D},\beta_{P},\beta_{D} \right)$.
If 
\begin{align*}
\lambda^{2} \gamma 
& \leq 
\frac{\epsilon^{3}}
 {\left(\norm{A}+3 \right)^{5} 
  \left(1 + \norm{\xxs} + \norm{\yys} \right)^{4}
 }\\
& \leq 
\left(            \frac{\epsilon}{\left(\norm{A}+2 \right)^{2} 
                      \left(\max \left(\norm{\xxs}, \norm{\yys} \right)+ 1 \right)}
 \right)^{2}
\left(   \frac{\epsilon }{\left(1 + \norm{\xxs}^{2} + \norm{\yys}^{2} \right)
                    \left(\norm{A} + 3 \right)}
 \right)
\end{align*}
then either
\[
  \lambda 
\leq 
            \frac{\epsilon}{\left(\norm{A}+2 \right)^{2} 
                      \left(\max \left(\norm{\xxs}, \norm{\yys} \right)+ 1 \right)}
\]
or
\[
  \gamma
\leq 
   \frac{\epsilon }{\left(1 + \norm{\xxs}^{2} + \norm{\yys}^{2} \right)
                    \left(\norm{A} + 3 \right)}.
\]
So,
\begin{align*}
& \prob{A,\bb ,\cc}
  {\lambda^{2} \gamma 
  \leq 
\frac{\epsilon^{3}}
 {\left(\norm{A}+3 \right)^{5} 
  \left(1 + \norm{\xxs} + \norm{\yys} \right)^{4}
 }
 \text{ and } \mathcal{F} (A, \bb ,\cc ) }\\
& \quad  \leq 
\frac{\epsilon n e }
     {\sigma^{2}} 
+ 
  \frac{8 \epsilon \left( (n + 1/2) (m+1) + (m + 1/2) (n+1) \right)}
       {\sigma^{2}}\\
& \quad \leq 
  \frac{16 \epsilon (n + 1) (m+1)}
       {\sigma^{2}}.
\end{align*}
As Lemma~\ref{lem:canRound} tells us that
\[
  \delta (A, \bb ,\cc )
\geq 
  \frac{
   \lambda^{2} \gamma 
   }{
   2 \max (1, \sqrt{n} \norm{A}) \left(1 + \norm{A} \right)
   },
\]
we obtain
\begin{align*}
& \prob{A,\bb ,\cc}
  {\delta 
  \leq 
\frac{\epsilon^{3}}
 {\left(\norm{A}+3 \right)^{5} 
  \left(1 + \norm{\xxs} + \norm{\yys} \right)^{4}
 }
\left(\frac{1}{2 \max (1,\sqrt{n}\norm{A}) \left(1 + \norm{A} \right)}
 \right)
  \text{ and } \mathcal{F} (A, \bb ,\cc )}\\
& \leq 
\prob{A,\bb ,\cc}
  {\lambda^{2} \gamma 
  \leq 
\frac{\epsilon^{3}}
   {
     \left(\norm{A}+3 \right)^{5} 
     \left(1 + \norm{\xxs} + \norm{\yys} \right)^{4}
   }  \text{ and } \mathcal{F} (A, \bb ,\cc )
} \\
& \leq 
  \frac{16 \epsilon (n + 1) (m+1)}
       {\sigma^{2}}.
\end{align*}
From this inequality, we derive
\begin{align*}
& \prob{A, \bb ,\cc }
 {\delta
 \left(\norm{A} + 3 \right)^{7}
  \left(1 + \norm{\xxs} + \norm{\yys } \right)^{4}
\leq 
\frac{\epsilon^{3}}
 {2 \sqrt{n}}
 \text{ and } \mathcal{F} (A, \bb ,\cc )
 }\\
& \leq  \prob{A,\bb ,\cc}
  {\delta 
\left(\norm{A}+3 \right)^{5} 
  \left(1 + \norm{\xxs} + \norm{\yys} \right)^{4}
 \max (1,\norm{A}) \left(1 + \norm{A} \right)
  \leq 
\frac{\epsilon^{3}}
 {2 \sqrt{n}}
 \text{ and } \mathcal{F} (A, \bb ,\cc )
 }\\
& \leq 
  \frac{16 \epsilon (n + 1) (m+1)}
       {\sigma^{2}}.
\end{align*}
The lemma now follows by changing
  $\epsilon^{3} / (2 \sqrt{n})$ to $\epsilon$.
\end{proof}

To convert this bound on the probability that
  $\delta$ is small to a bound on
  the expectation of $\log (1/\delta)$, we will
  use the following technical lemma:

\begin{lemma}\label{lem:toLog}
Let $x$ be a non-negative random variable for which
  there exist constants $\alpha$ and $k$ such that
  $\log (\alpha)/ k \geq 1$  and
\[
  \prob{}{x \leq \epsilon }
\leq 
  \alpha \epsilon ^{k}.
\]
Then,
\[
  \expec{}{\max \left(1, \log (1/x) \right)}
 \leq 
  \frac{1 + \log \alpha}{k}.
\]
\end{lemma}
\begin{proof}
We compute
\begin{align*}
  \expec{}{\max \left(1,\log (1/x) \right)}
& =
  \int_{t=0}^{\infty} \prob{}{\max \left(1,\log (1/x) \right) \geq t} \diff{t}\\
& =
  \int_{t=0}^{\frac{\log \alpha}{k}} \prob{}{\max \left(1,\log (1/x) \right) \geq t} 
           \diff{t}
+
  \int_{\frac{\log \alpha}{k}}^{\infty } \prob{}{\max \left(1,\log (1/x) \right) \geq t} 
           \diff{t}\\
& \leq 
  \int_{t=0}^{\frac{\log \alpha}{k}} 
           \diff{t}
+
  \int_{\frac{\log \alpha}{k}}^{\infty } \prob{}{\log (1/x) \geq t} 
           \diff{t},
\qquad \text{ as $(\log \alpha)/ k \geq 1$,}\\
& \leq 
  \int_{t=0}^{\frac{\log \alpha}{k}} 
           \diff{t}
+
  \int_{\frac{\log \alpha}{k}}^{\infty }
      \alpha e^{-t k}
           \diff{t}\\
& =
  \frac{\log \alpha}{k}
+
  \frac{1}{k}.
\end{align*}
\end{proof}

From this, we obtain
\begin{corollary}\label{cor:logDelta}
Under the conditions of Theorem~\ref{thm:delta},
\begin{multline*}
\expec{}{
\max \left(1,
\log \left(\frac{1}{\delta (A, \bb ,\cc )} \right) 
\right)
}\\
\leq 
3 (\log (21 (m +1)^{13/6} / \sigma^{2}) + 1)
+ 7 \log (\norm{A} + 3)
+ 4 \log (1 + \norm{\xxs} + \norm{\yys })
\end{multline*}
\end{corollary}
\begin{proof}
Applying Lemma~\ref{lem:toLog} to \eqref{eqn:reduce},
  and recalling $m \geq n$,
  we obtain
\begin{multline*}
\expec{}{
\max \left(1,
\log \left(\frac{1}{\delta (A, \bb ,\cc )} \right)
 \right)
}
- 7 \expec{}{\log (\norm{A} + 3)}
- 4 \expec{}{\log (1 + \norm{\xxs} + \norm{\yys })}\\
\leq 
3 \left(\log (21 (m+1)^{13/6} / \sigma^{2}) + 1 \right) .
\end{multline*}
\end{proof}

As $\norm{\xxs}$ and $\norm{\yys}$ can be bounded in
  terms of the condition number of the linear program,
  we will be able to use Theorem~\ref{thm:DST} 
  to bound the probability that they are large.
The probability that $\norm{A}$ is large may be obtained
  by more elementary means.
In particular, we prove

To bound $\expec{}{\log (1 + \norm{\xxs } + \norm{\yys })}$, we note
  that Renegar~\cite[Propositions 2.2 and 2.3]{RenegarCond} has proved
\begin{lemma}[Norms of optimal solutions]\label{lem:renegarNorm}
\[
 \max \left( \norm{\xxs}, \norm{\yys} \right) \leq C (A,\bb ,\cc)^{2}.
\]
\end{lemma}
So, we may apply Theorem~\ref{thm:DST} to bound the norms of $\xxs$ and $\yys$.
To bound the norm of $A$, we apply:

\begin{proposition}\label{pro:expecNorm}
Let $A$ be a Gaussian perturbation of variance 
  $\sigma^{2} \leq 1$ of an $m$-by-$n$
  matrix $\orig{A}$ of norm at most $1$.
Then,
\[
  \expec{}{\log (\norm{A} + 3)} 
  \leq \log (\left(\sqrt{n} + \sqrt{m} \right) \sigma  + 4)
\]
\end{proposition}
\begin{proof}
Write $A = \orig{A} + G \sigma $ where
  $G$ is a Gaussian random matrix of variance
  $1$ centered at the origin and $\norm{\orig{A}} \leq 1$.
Seginer~\cite{Seginer} proves that
  $\expec{}{\norm{G}} \leq \sqrt{n} + \sqrt{m}$,
  which implies
  $\expec{}{\norm{G \sigma }} \leq \sigma (\sqrt{n} + \sqrt{m})$
  and
\[
  \expec{}{\norm{A} + 3} 
  \leq (\left(\sqrt{n} + \sqrt{m} \right) \sigma  + 4).
\]
As the logarithm is a convex function,
   
\[
  \expec{}{\log (\norm{A} + 3)} 
\leq 
  \log \left(\expec{}{\norm{A} + 3} \right) 
  \leq \log \left(\left(\sqrt{n} + \sqrt{m} \right) \sigma  + 4 \right).
\]
\end{proof}

Putting this all together, we prove the main theorem:

\begin{proof}[Proof of Theorem~\ref{thm:delta}]
To bound the terms obtained in Corollary~\ref{cor:logDelta},
  we apply Proposition~\ref{pro:expecNorm} to show
\[
  \expec{}{\log (\norm{A} + 3)} \leq  \log (\sqrt{n} + \sqrt{m} + 4).
\]
We then apply Lemma~\ref{lem:renegarNorm} to show
\[
  (1 + \norm{\xxs} + \norm{\yys })
\leq 
  3 C (A,\bb ,\cc)^{2},
\]
and Theorem~\ref{thm:DST} to obtain
\begin{align*}
  \expec{}{\log   (1 + \norm{\xxs} + \norm{\yys })}
& \leq 
  \expec{}{\log \left( 3 C (A,\bb ,\cc)^{2} \right)}\\
& \leq 
  \log (3) + 2 \expec{}{\log C (A,\bb ,\cc)}\\
& \leq 
  O (\log (mn/\sigma )).
\end{align*}
\end{proof}

\section{Geometric Analysis of $\delta $}\label{sec:geometric}

To prove Lemma~\ref{lem:canRound},
  we use the following lemma which says that if
  the value of $\cc \xx$ is close to optimal,
  then $\xx$ must be close to $\xxs$.

\begin{lemma}\label{lem:close}
For a linear program specified by $(A, \bb ,\cc)$ 
  with unique optimal primal and dual solutions $\xxs$ and $\yys$, 
  let 
\[
\lambda (A, \bb ,\cc ) = \min \left(\alpha_{P} (A,\bb ,\cc ),
	\alpha_{D} (A,\bb ,\cc ),
	\beta_{P} (A,\bb ,\cc ),
	\beta_{D} (A,\bb ,\cc ) \right).
\]
Then,
\[
  \norm{\xxs -\xx }_{\infty}
  \leq 
  \cc (\xxs - \xx) 
  \left(\frac{
       1 + \norm{A}
     }{
       \lambda (A,\bb ,\cc ) \min (\gamma (A, \bb ,\cc ) ,1)
     } \right)
\]  
\end{lemma}

\begin{proof}[Proof of Lemma~\ref{lem:canRound}]
Assuming
\[
\cc (\xxs -\xx) <
  \frac{
   \lambda (A ,\bb ,\cc )^{2} \gamma (A, \bb , \cc )
   }{
   2 \max (1,\sqrt{n} \norm{A}) \left(1 + \norm{A} \right)
   },
\]
we need to show $U (\xx ) = U$ and $V (\xx) = V$.
From Lemma~\ref{lem:close}, we have
\[
  \infnorm{\xxs -\xx } 
<
  \frac{
   \lambda ^{2} \gamma
   }{
   2 \max (1,\sqrt{n} \norm{A}) \left(1 + \norm{A} \right)
   }
  \left(\frac{
       1 + \norm{A}
     }{
       \lambda  \min (\gamma ,1)
     } \right)
\leq 
 \frac{\lambda }{2 \max (1,\sqrt{n} \norm{A})}
\leq  \frac{\lambda }{2}.
\]

We then have
\begin{itemize}
\item [(a)] for  $i \in U$, $x_{i} > 
  \alpha_{P} (A,\bb ,\cc) - \lambda  /2
  > \lambda  /2 $ ,
\item [(b)] for $i \not \in U$, $x_{i} <\lambda  /2$,
\end{itemize}
As $\abs{A_{j,:} (\xxs -\xx)} \leq  \infnorm{A} \infnorm{\xxs -\xx}
  \leq \sqrt{n} \norm{A} \infnorm{\xxs -\xx}$,
  we also have
\begin{itemize}
\item [(c)] for $j \in V$, $b_{j} - A_{j,:} \xx < \lambda /2$,
\item [(d)] for $j \not \in V$, 
  $b_{j} - A_{j,:} \xx 
   > \beta_{P} (A,\bb ,\cc) - \lambda  /2 
   > \lambda  /2$.
\end{itemize}
So, the smallest $n$ values in 
$\{\xx_{i}\} \cup \{\bb_{j} - A_{j,:}\xx\}$
are those indexed by $\Ub$ and $V$.
\end{proof}

The proof of Lemma~\ref{lem:close}  relies on the following
  technical lemmas.

\begin{lemma}\label{lem:techAv}
For $\xx$ a feasible point for a bounded linear program
  specified by $(A,\bb ,\cc )$,
\[
  \cc (\xxs -\xx) \geq \alpha_{D} (A,\bb ,\cc ) \norm{A_{V,:} (\xxs -\xx)}.
\]
\end{lemma}
\begin{proof}
As $\yys A \geq \cc$, we have
\begin{align*}
  \cc  (\xxs -\xx)
& \geq 
  \cc \xxs - \yys A \xx \\
& =   
  \yys A \xxs - \yys A \xx  & \text{ (by strong duality)}\\
& =
  \yys A (\xxs - \xx )\\
& =
  \yys_{V} A_{V,:} (\xxs - \xx ),\\
\end{align*}
as $\yys$ is zero outside of $V$.
As $A_{V,:} (\xxs - \xx )$ is non-negative,
 we may conclude that
\[
  \yys_{V} A_{V,:} (\xxs - \xx )
 \geq 
  \alpha_{D} (A,\bb ,\cc ) \norm{ A_{V,:} (\xxs - \xx )}_{1}
 \geq 
  \alpha_{D} (A,\bb ,\cc ) \norm{ A_{V,:} (\xxs - \xx )}.
\]
\end{proof}

\begin{lemma}\label{lem:techGamma}
For $\xx$ a feasible point for a bounded linear program
  specified by $(A,\bb ,\cc )$,
\[
   \norm{A_{V,U} (\xxs_{U} - \xx_{U})}
 \geq 
   \gamma (A, \bb , \cc )\infnorm{\xxs_{U} - \xx_{U}}.
\]
\end{lemma}
\begin{proof}
For any  $k \in U$, 
 let $\qq$ be the null vector of the span of $A_{V,U \setminus k}$.
Then,
\[
  \qq A_{V,U} \xx 
= 
  \qq \left( A_{V,k} x_{k} + A_{V,U-k} \xx_{U-k} \right)
=
x_{k} \qq  A_{V,k}
=
x_{k} \dist{A_{V,k}}{\Span{A_{V , U \setminus k}}}.
\]
So,
\begin{align*}
 \norm{A_{V,U} (\xxs_{U} - \xx_{U})}
& \geq 
 \abs{ \qq A_{V,U}  (\xxs_{U} - \xx_{U})} \\
& =
 \abs{\xs_{k} - x_{k}}
    \dist{A_{V,k}}{\Span{A_{V , U \setminus k}}}\\
& \geq 
 \abs{\xs_{k} - x_{k}}  \gamma (A, \bb , \cc ).
\end{align*}
\end{proof}

\begin{lemma}\label{lem:techUbar}
For $\xx$ a feasible solution to a linear program 
  specified by $(A , \bb ,\cc )$,
\[
 \norm{\xx_{\Ub}} 
\leq 
 \frac{\cc (\xxs -\xx)}{\beta_{D} (A,\bb ,\cc )}.
\]
\end{lemma}

\begin{proof}
As $\yys A_{:,U} = \cc_{U}$, 
 $(\yys A-\cc )\xx = (\yys A_{:, \Ub}-\cc_{\Ub})\xx_{\Ub} $.

As
  every entry in
  $\yys A_{:, \Ub} - \cc_{\Ub} $ 
  is at least
  $\beta_{D} (A, \bb ,\cc )$
  and $\xx \geq 0$,  we have
\begin{equation}\label{eqn:techUbar}
(\yys A-\cc )\xx = (\yys A_{:,\Ub}-\cc_{\Ub})\xx_{\Ub}
  \geq \beta_{D} (A,\bb ,\cc ) \onenorm{\xx_{\Ub}}
  \geq \beta_{D} (A,\bb ,\cc ) \norm{\xx_{\Ub}}.
\end{equation}
As $\yys \geq 0$,
\[
\cc \xxs -\cc \xx 
 =\yys \bb -\cc \xx
 \geq \yys A\xx -\cc \xx 
 \geq \beta_{D} (A,\bb ,\cc ) \norm{\xx_{\Ub}},
\]
where the last inequality follows from \eqref{eqn:techUbar}.
\end{proof}

\begin{proof}[Proof of Lemma~\ref{lem:close}]
Applying the triangle inequality, we observe
\[
  \norm{A_{V,:} (\xxs -\xx )}
\geq 
  \norm{A_{V,U} (\xxs_{U} -\xx_{U} )}
-
  \norm{A_{V,\Ub} (\xxs_{\Ub} -\xx_{\Ub} )}.
\]
We can bound the first of these terms by
  applying Lemma~\ref{lem:techGamma}, and
  the second by observing
  $\xxs _{\Ub} = 0$ and
  $\norm{A_{V,\Ub} \xx_{\Ub}} 
 \leq \norm{A_{V,\Ub}} \norm{\xx _{\Ub}}$,
thereby proving
\[
  \norm{A_{V,:} (\xxs -\xx )}
\geq 
  \gamma (A, \bb ,\cc ) \infnorm{\xxs _{U} - \xx _{U}}
-
  \norm{A_{V,\Ub}} \norm{\xx _{\Ub}}.
\]
By now applying Lemma~\ref{lem:techAv}, we obtain
\begin{align*}
  \cc  (\xxs -\xx )
  \geq 
\alpha_{D} (A, \bb ,\cc ) \left(  \gamma (A, \bb ,\cc )\infnorm{\xxs _{U} - \xx _{U}}
-
  \norm{A_{V,\Ub}} \norm{\xx _{\Ub}}
 \right),
& \text{ which implies}\\
\alpha_{D} (A, \bb ,\cc ) \norm{A_{V,\Ub}} \norm{\xx _{\Ub}}
+  \cc  (\xxs -\xx )
  \geq 
\alpha_{D} (A, \bb ,\cc ) \gamma (A,\bb ,\cc ) \infnorm{\xxs _{U} - \xx _{U}}.
\end{align*}
As Lemma~\ref{lem:techUbar} implies
  $\cc (\xxs -\xx) \geq \beta_{D} (A,\bb ,\cc ) \norm{\xx _{\Ub}}$
  and $\norm{A} \geq \norm{A_{V,\Ub }}$,
  we obtain
\begin{align*}
\left(1 + \frac{\alpha_{D} (A,\bb ,\cc )}{\beta_{D} (A,\bb ,\cc )}
          \norm{A} \right)
  \cc  (\xxs -\xx )
  \geq 
\alpha_{D} (A,\bb ,\cc ) \gamma (A,\bb ,\cc ) \infnorm{\xxs _{U} - \xx _{U}}
& \text{ which implies}\\
\frac{  \cc  (\xxs -\xx ) \left(1 + \norm{A} \right)
}{
\min \left(\alpha_{D} (A,\bb ,\cc ),\beta_{D} (A,\bb ,\cc )  \right)
}
  \geq 
\gamma (A, \bb ,\cc) \infnorm{\xxs _{U} - \xx _{U}}.
\end{align*}
The lemma now follows from this inequality and
  Lemma~\ref{lem:techUbar}, which implies
\[
\frac{\cc  (\xxs - \xx)}{\beta_{D} (A,\bb ,\cc )}
\geq 
\norm{\xx _{\Ub }}
\geq 
\infnorm{\xx _{\Ub }}
=
\infnorm{\xxs_{\Ub} - \xx_{\Ub }}.
\]
\end{proof}

\section{Bounds on $\alpha$, $\beta$, and $\gamma $}\label{sec:probabilistic}
For this section, we let $\mu_{A}$, $\mu_{\bb}$ and $\mu_{\cc}$  
  denote the Gaussian densities on $A$, $\bb$ and $\cc$ in Theorem~\ref{thm:delta}.
For an index $j$ or set of indices $V$, we let
  $\mu_{b_{j}}$ and $\mu_{\bb_{V}}$ denote the induced distributions
  on $b_{j}$ and $\bb_{V}$, and we extend this notational convention
  to sub-matrices of $A$ and sub-vectors of $\cc$.

The idea behind our proofs 
  of Lemma~\ref{lem:probAlpha}, \ref{lem:probBeta} and \ref{lem:probGamma} 
  is that for any configuration
  of $A$, $\bb$ and $\cc$ in which $\alpha$, $\beta$, or $\gamma$
  is small, there are many nearby configurations
  in which the term is not too small.
As Gaussian densities do not fall off too quickly,
  this nearby configuration will have approximately
  the same probability as the original.
To make this idea rigorous, we establish mappings
  pairing configurations in which these terms are
  small with configurations in which these terms are not.
We then use these mappings to 
  show that the Gaussian probability of the configurations
  in which the terms are not small is much larger
  than those in which they are.

To show that it is unlikely that $\beta_{P}$ is small,
  we hold $A$, $\yys$, $\cc$, and $\xxs$ constant,
  and map those $\bb_{j}$'s that are close to
  $A_{j,:} \xxs$  to be a little further away.
To show that it is unlikely that $\alpha_{P}$ is small,
  we hold $A$, $\yys$ and $\cc$ constant, and map
  small non-zero entries of $\xxs$ to larger values
  while simultaneously mapping the entries of $\bb$ 
  to preserve the tight constraints
  and maintain slack in the others.
To show that it is unlikely that $\gamma$ is small,
  we hold $\xxs$, $\yys$, and the slack components
  of $\bb$ and $\cc$ constant.
We then vary $A_{V,U}$ slightly,
  changing
  $\bb_{V}$ and $\cc_{U}$ accordingly.
As each slight motion described only induces a slight motion
  in the other components, we can prove
  that each configuration obtained has similar
  probability.

To turn these intuitive arguments into proofs, we need
  four tools:
\begin{enumerate}
\item [1.] a bound on the smoothness of the Gaussian density,
\item [2.] a bound on the probability that a random variable
  is small given that its density is smooth near zero,
\item [3.] a lemma making rigorous the change of variables implicitly
  used in the intuitive arguments, and
\item [4.] a proof that the probability of an event
  can be bounded by the maximum of its probability
  over the sets in a partition of its probability space.
\end{enumerate}
Each of these tools is relatively simple, and the last
  should be obvious for finite partitions.
The bound on the smoothed complexity of the 
  simplex method~\cite{SpielmanTengSimplex} uses
  each of these tools along with some others.
It is our hope that the reader would 
  have an easier time understanding the proofs 
  in~\cite{SpielmanTengSimplex} after having read
  this section.

We now develop these four tools, and at the end of the section
  apply them to the proofs of the bounds on $\alpha$, $\beta$ and $\gamma$.

We make use of the following elementary bound on
  the smoothness of Gaussians:
\begin{lemma}[Smoothness of Gaussians]\label{lem:GaussianSmooth}
Let $\mu (\xx)$ be a Gaussian distribution
  in $\Reals{n}$ 
  of variance $\sigma ^{2}$
  centered at a point of norm at most 1.
If $\dist{\xx}{\yy} < \epsilon \leq 1$, then
\[
  \frac{\mu (\yy)}{\mu (\xx )} 
  \geq 
  e^{- \frac{\epsilon (\norm{\xx} + 2)}{\sigma^{2}}}.
\]
\end{lemma}
\begin{proof}
Let $\orig{\xx}$ be the center of the distribution.
We compute
\begin{align*}
  \frac{\mu (\yy)}{\mu (\xx )} 
& =
 e^{\frac{-1}{2 \sigma^{2}} 
  \left(\norm{\yy - \orig{\xx}}^{2} - 
        \norm{\xx - \orig{\xx}}^{2} \right)}\\
& \geq 
 e^{\frac{-1}{2 \sigma^{2}} 
  \left(\left(\norm{\yy - \xx} + 
              \norm{\xx - \orig{\xx}} \right)^{2} - 
        \norm{\xx - \orig{\xx}}^{2} \right)}
& \text{by the triangle inequality}\\
& =
 e^{\frac{-1}{2 \sigma^{2}} 
  \left(2 \norm{\yy - \xx} \norm{\xx - \orig{\xx}} 
        + \norm{\yy - \xx}^{2} \right)}\\
& \geq 
 e^{\frac{-1}{2 \sigma^{2}} 
  \left(2 \epsilon  \norm{\xx - \orig{\xx}}
        + \epsilon ^{2} \right)}\\
& \geq 
 e^{\frac{-1}{2 \sigma^{2}} 
  \left(2 \epsilon \left( \norm{\xx} + 1 \right)
        + \epsilon ^{2} \right)}
& \text{as $\norm{\orig{\xx}} \leq 1$}\\
& \geq 
 e^{\frac{-1}{2 \sigma^{2}} 
  \left(2 \epsilon \left( \norm{\xx} + 1 \right)
        + \epsilon  \right)}
  & \text{ as $\epsilon \leq 1$}\\
& \geq 
 e^{-\frac{\epsilon \left( \norm{\xx} + 2 \right)}{ \sigma^{2}} }.
\end{align*}
\end{proof}

We remark that this lemma is the only fact about Gaussian random
  variables used in this paper.
Thus, one could obtain results of a similar character for
  any distribution that satisfies properties similar
  to those derived for Gaussian random vectors above.

The argument by which we obtain probability bounds
  from comparing configurations 
  is encapsulated in the following lemma,
  which is used in each of the three proofs.
This lemma essentially says that if a distribution
  of a random variable is relatively flat near a point, then
  the variable is unlikely to lie too close to that point.

\begin{lemma}[Smooth distributions unlikely small]\label{lem:ratio}
Let $x$ be a real random variable distributed according
  to density $\rho$ such that there exist constants
  $\alpha$ and $c$ for which
\[
\text{$0 \leq x \leq x' \leq \alpha $ implies }
\quad 
  \frac{\rho (x')}{\rho (x)} \geq c.
\]
Then, for $\epsilon < \alpha $,
\[
\prob{}{x \in [0,\epsilon ] \big| x \in [0,\alpha ]}
\leq \frac{\epsilon}{c \alpha }.
\]
In particular, 
\[
\prob{}{x \in [0,\epsilon ]}
 \leq \frac{\epsilon}{c \alpha }.
\]
\end{lemma}
\begin{proof}
From the definition of conditional probability,
  we have
\[
\prob{}{x \in [0,\epsilon ] \big| x \in [0,\alpha ]}
=
\frac{
  \int_{0}^{\epsilon} \rho (x) \diff{x}
}{
  \int_{0}^{\alpha } \rho (x) \diff{x}
}.
\]
Setting $y = (\epsilon /\alpha)x$, we compute
\[
  \int_{0}^{\alpha}\rho (x) \diff{x}
 = 
(\alpha /\epsilon) 
  \int_{0}^{\epsilon}\rho ((\alpha /\epsilon)y) \diff{y}
\geq 
(\alpha /\epsilon) 
  \int_{0}^{\epsilon} c \rho (y) \diff{y}.
\]
From which the lemma follows.
\end{proof}

For example, we can use the previous lemma to derive a bound
  on the probability that a Gaussian random variable is 
  greater than $t+\epsilon$ given that it is greater than $t$:

\begin{lemma}[Comparison of Gaussian tails]\label{lem:GaussianTails}
Let $x$ be a Gaussian random variable of variance $\sigma^{2} \leq 1$
  and mean of absolute value at most $1$.
For $\epsilon \geq 0$, $\tau \geq 1$ and $t \leq  \tau $,
\[
\prob{}{x \leq t + \epsilon \big| x \geq t}
\leq 
\frac{\epsilon \tau}{\sigma^{2}}
e^{\frac{\epsilon (\tau +3)}{\sigma^{2}}}
\quad \text{and} \quad 
\prob{}{x \geq t + \epsilon \big| x \geq t}
\geq 
1-
\frac{\epsilon \tau}{\sigma^{2}}
e^{\frac{\epsilon (\tau +3)}{\sigma^{2}}}.
\]
\end{lemma}
\begin{proof}
It suffices to prove the first bound.
Let $\mu$ be the density function of $x$.
Let $\alpha = \sigma^{2} / \tau \leq 1$.
For $\epsilon \geq  \alpha$ the lemma is vacuous.
For $\epsilon < \alpha$, we will show that
\begin{equation}\label{eqn:tails}
\text{$  t \leq x < x' \leq t + \alpha$ implies }
\frac{\mu (x')}{\mu (x)}
\geq 
e^{- \frac{\epsilon (\tau +3)}{\sigma^{2}}},
\end{equation}
and then apply Lemma~\ref{lem:ratio} to finish the proof.
For $t < -1$, \eqref{eqn:tails} is trivial as
  $\mu$ is monotone increasing
  on $[t, t + \alpha]$.
For $t \geq -1$, we have $\norm{x} \leq \tau +1$
  so \eqref{eqn:tails} follows from
  Lemma~\ref{lem:GaussianSmooth}.
\end{proof}

Finally, we note that our intuitive explanation of
  the proofs of Lemmas~\ref{lem:probAlpha},
 \ref{lem:probBeta} and \ref{lem:probGamma} 
  implicitly used a change of variables:
  instead of reasoning in terms of the variables
  $A$, $\bb$ and $\cc$, we found it more convenient
  to think of $\xxs$ and $\yys$ as quantities to
  fix or vary. 
We now introduce the machinery that enables us
  to reason in terms of these variables.
We begin by observing that for any sets $U$ and $V$,
  not necessarily the combinatorial type of
  $(A, \bb , \cc )$, we can
  introduce variables $\xxs_{U}$ and $\yys_{V}$,
  not necessarily the optimal primal and dual solutions,
  and define
\[
   \bb_{V} = A_{V,U} \xxs_{U}
\quad \text{ and} \quad 
   \cc_{U} = \yys_{V} A_{V,U}.
\]
We can then compute probabilities in these
  new variables by observing that the
  joint density of
  $A$, $\xxs_{U}$, $\yys_{V}$, $\bb_{\Vb}$, and
  $\cc_{\Ub}$ is
\[
   \ma{A} \mbi{A_{V,U} \xxs_{U}} \mci{\yys_{V} A_{V,U}}
   \mbo{\bb_{\Vb}} \mco{\cc_{\Ub}} \det{A_{V,U}}^{2}.
\]
To see why this is true, recall that probabilities
  are best understood as integrals, and that the
  probability of an event $\mathcal{E} (A, \bb , \cc )$
  is
\begin{equation}\label{eqn:integral}
 \int_{A, \bb , \cc} \ind{\mathcal{E} (A, \bb ,\cc )}
  \ma{A} \mb{\bb} \mc{\cc } 
  \diff{A} \diff{\bb} \diff{\cc} 
\end{equation}
To express this integral in the new variables, we first
  compute the Jacobian of the
  change of variables, which is
\[
\abs{\det{\frac{\partial (A, \bb_{\Vb}, \cc_{\Ub}, \bb_{V}, \cc_{U})}
  {\partial (A, \bb_{\Vb}, \cc_{\Ub}, \xxs_{U}, \yys_{V})}}}
= \det{A_{V,U}}^{2};
\]
so,
\begin{align*}
  \diff{A} \diff{\bb} \diff{\cc}
& = 
\abs{\det{\frac{\partial (A, \bb_{\Vb}, \cc_{\Ub}, \bb_{V}, \cc_{U})}
  {\partial (A, \bb_{\Vb}, \cc_{\Ub}, \xxs_{U}, \yys_{V})}}}
\diff{A} \diff{\bb_{\Vb }}  \diff{\cc_{\Ub}} 
  \diff{\xxs_{U}}  \diff{\yys_{V}}
\\
& =
\det{A_{V,U}}^{2}
\diff{A} \diff{\bb_{\Vb }}  \diff{\cc_{\Ub}} 
  \diff{\xxs_{U}}  \diff{\yys_{V}},
\end{align*}
and
\begin{align*}
\eqref{eqn:integral} =
\int_{A, \bb_{\Vb } , \cc_{\Ub}, \xxs_{U} ,\yys_{V}}
&  \ind{\mathcal{E} (A, \bb ,\cc )}
   \ma{A} \mbi{A_{V,U} \xxs_{U}} \mci{\yys_{V} A_{V,U}}
   \mbo{\bb_{\Vb}} \mco{\cc_{\Ub}} \det{A_{V,U}}^{2}\\
& \cdot \diff{A} \diff{\bb_{\Vb }}  \diff{\cc_{\Ub}} 
  \diff{\xxs_{U}}  \diff{\yys_{V}}.
\end{align*}
While we can define this change of variables for
  any sets $U$ and $V$, we will of course only apply
  this change of variables to programs of type $(U,V)$.
If we let $\typeuv$ denote the set of
  $(A, \bb, \cc)$ of type $(U,V)$, then
  we can express the probability of
  $\ind{\mathcal{E} (A, \bb ,\cc ) \mbox{ and } \mathcal{F} (A, \bb ,\cc )}$ as
\begin{multline*}
  \int_{A, \bb , \cc : \mathcal{F} (A, \bb ,\cc )} 
   \ind{\mathcal{E} (A, \bb ,\cc )}
  \ma{A} \mb{\bb} \mc{\cc }
  \diff{A} \diff{\bb} \diff{\cc}\\
=
\sum_{U,V}
  \int_{A, \bb , \cc : \typeuv } 
   \ind{\mathcal{E} (A, \bb ,\cc )}
  \ma{A} \mb{\bb} \mc{\cc }
  \diff{A} \diff{\bb} \diff{\cc},
\end{multline*}
and then apply the change of variables corresponding
  to $(U,V)$ to evaluate the integral
  over $\typeuv$ on the right.
In fact, in each of our proofs,
  we will actually bound
\[
  \max_{U,V} \prob{}{\mathcal{E} (A, \bb ,\cc ) | \typeuv}.
\]
To see that this upper bounds the probability
  of $\ind{\mathcal{E} (A, \bb ,\cc ) \mbox{ and } \mathcal{F} (A, \bb ,\cc )}$, we prove
\begin{claim}\label{clm:upperBound}
\[
\prob{}{\mathcal{E} (A, \bb ,\cc ) \text{ and } \mathcal{F} (A, \bb ,\cc)}
 \leq 
  \max_{U,V} \prob{}{\mathcal{E} (A, \bb ,\cc ) | \typeuv}.
\]
\end{claim}
\begin{proof}
\begin{align*}
\prob{}{\mathcal{E} (A, \bb ,\cc )\text{ and } \mathcal{F} (A, \bb ,\cc)}
& = 
\sum_{U,V}
\prob{}{\mathcal{E} (A, \bb ,\cc ) \text{ and } \typeuv }
\\
& =
\sum_{U,V}
\prob{}{\typeuv }
\prob{}{\mathcal{E} (A, \bb ,\cc ) | \typeuv }\\
& \leq 
\max_{U,V}
\prob{}{\mathcal{E} (A, \bb ,\cc ) | \typeuv },
\end{align*}
as $\sum_{U,V} \prob{}{\typeuv } \leq 1$.
\end{proof}

We summarize this discussion in the following lemma:

\begin{lemma}[Change of variables]\label{lem:cov}
Let $\mathcal{E} (A, \bb ,\cc )$ be an event.
Then,
\begin{multline*}
  \prob{A, \bb ,\cc}{\mathcal{E} (A, \bb ,\cc ) \text{ and } \mathcal{F} (A ,\bb ,\cc )}\\
\leq 
\max_{U,V}
\prob{A, \xxs ,\yys ,\bb_{\Vb}, \cc_{\Ub }}
     {\mathcal{E} (A, \bb ,\cc ) | \text{$A_{\Vb ,:} \xxs \leq \bb_{\Vb}$
                 and
               $\yys  A_{:,\Ub } \geq \cc_{\Ub}$}},
\end{multline*}
where $A$, $\xxs$, $\yys$, $\bb_{\Vb}$ and $\cc_{\Ub}$
  have joint density
\[ 
   \ma{A} \mbi{A_{V,U} \xxs_{U}} \mci{\yys_{V} A_{V,U}}
   \mbo{\bb_{\Vb}} \mco{\cc_{\Ub}} \det{A_{V,U}}^{2}.
\]
\end{lemma}

In fact, all of our proofs begin by fixing some
  subset of the variables, and then proving a probability
  bound for any configuration of the fixed variables.
This amounts to proving a probability upper bound by
  dividing the probability space into a number of regions,
  and proving that the bound holds in each of these regions.
Formally, we are using the fact:

\begin{proposition}[Upper bound by max of probabilities]\label{pro:max}
Let $X$ and $Y$ be random variables
 distributed according to an integrable density
  function $\mu (X,Y)$
 and let
  $\mathcal{E} (X,Y)$ be an event.
Then
\[
  \prob{X,Y}{\mathcal{E} (X,Y)}
\leq 
  \max_{y} \prob{X,Y}{\mathcal{E} (X,Y) | Y = y}
\defeq 
  \max_{Y} \prob{X}{\mathcal{E} (X,Y) | Y }.
\]
\end{proposition}
\begin{proof}
By Tonelli's Theorem, we have
\begin{align*}
  \prob{X,Y}{\mathcal{E} (X,Y)}
& = 
\int_{X,Y} \ind{\mathcal{E} (X,Y)} \mu (X,Y) \diff{X} \diff{Y}\\
& =
\int_{Y} \left(
 \int_{X} \ind{\mathcal{E} (X,Y)} \mu (X,Y) \diff{X} \right)  \diff{Y}\\
& =
\int_{Y} 
 \left(\int_{X}  \mu (X,Y) \diff{X} \right)
 \left(
\frac{
 \int_{X} \ind{\mathcal{E} (X,Y)} \mu (X,Y) \diff{X}
}{
 \left(\int_{X}  \mu (X,Y) \diff{X} \right)
}
 \right)  \diff{Y}\\
& =
\int_{Y} 
 \left(\int_{X}  \mu (X,Y) \diff{X} \right)
 \left(
\prob{X}{\mathcal{E} (X,Y) | Y}
 \right)  \diff{Y}\\
& \leq 
\max_{Y} 
\prob{X}{\mathcal{E} (X,Y) | Y},
\end{align*}
as
\[
\int_{Y} 
 \left(\int_{X}  \mu (X,Y) \diff{X} \right) \diff{Y} = 1.
\]
\end{proof}

Having established these tools, we now proceed
  with the proofs of Lemmas~\ref{lem:probAlpha}, \ref{lem:probBeta}
  and \ref{lem:probGamma}.

\begin{proof}[Proof of Lemma~\ref{lem:probBeta}
  (Probability of small $\beta $)]
By Lemma~\ref{lem:cov}, it suffices to bound
\[
\max_{U,V}
\prob{A, \xxs ,\yys ,\bb_{\Vb}, \cc_{\Ub }}
     {\beta_{P} (A, \bb ,\cc) 
    \leq  \frac{\epsilon }{\max \left(1,\norm{A} \norm{\xxs} \right)}
   \Big| \text{$A_{\Vb ,:} \xxs \leq \bb_{\Vb}$
                 and
               $\yys  A_{:,\Ub } \geq \cc_{\Ub}$}}.
\]
By Proposition~\ref{pro:max}, it suffices to
  prove that for all
  $U$, $V$, $A$, $\xxs_{U}$, $\yys_{V}$ and $\cc_{\Ub }$,
\begin{align*}
&  \prob{\bb_{\Vb}}
       {\exists j \in \Vb: b_{j} - A_{j,:} \xxs \leq
             \epsilon'
       \quad  \big| \quad 
  \forall j :  b_{j} - A_{j,:} \xxs  \ge 0}\\
& \leq 
\sum_{j \in \Vb }  \prob{\bb_{j}}
       {b_{j} - A_{j,:} \xxs \leq
             \epsilon'
       \quad  \big| \quad 
  \forall j :  b_{j} - A_{j,:} \xxs  \ge 0}\\
& =
\sum_{j \in \Vb }  \prob{\bb_{j}}
       {b_{j} - A_{j,:} \xxs \leq
             \epsilon'
       \quad  \big| \quad 
   b_{j} - A_{j,:} \xxs  \ge 0}\\
& =
\sum_{j \in \Vb }  \prob{\bb_{j}}
       {b_{j}  \leq  A_{j,:} \xxs +
             \epsilon'
       \quad  \big| \quad 
   b_{j}  \ge  A_{j,:} \xxs}\\
& \leq 
\frac{m \epsilon' \left(\norm{A} \norm{\xxs}\right)}
     {\sigma^{2}}
e^{
\frac{\epsilon' \left(\norm{A} \norm{\xxs} + 3\right)}
     {\sigma^{2}}
} 
& \text{by Lemma~\ref{lem:GaussianTails}}.\\
\end{align*}
Setting $\epsilon = \epsilon' \max \left(1, \norm{A} \norm{\xxs } \right)$,
  and observing that the lemma is vacuously true for 
 $\epsilon > \sigma^{2}/4m$, we deduce
\[
 \prob{\bb_{\Vb}}
       {\exists j \in \Vb: b_{j} - A_{j,:} \xxs \leq
             \epsilon
       \quad  \big| \quad 
  \forall j :  b_{j} - A_{j,:} \xxs  \ge 0}
\leq 
\frac{m \epsilon}
     {\sigma^{2}}
e^{
\frac{4 \epsilon}
     {\sigma^{2}}
}
\leq 
\frac{e m \epsilon}
     {\sigma^{2}}
\leq 
\frac{4 m \epsilon}
     {\sigma^{2}},
\]
for $\epsilon < \sigma^{2} / 4m$.
\end{proof}

\begin{proof}[Proof of Lemma~\ref{lem:probGamma} 
(Probability of small $\gamma$)]
By Lemma~\ref{lem:cov}, it suffices to bound
\[
\max_{U,V}
\prob{A, \xxs_{U} ,\yys_{V} ,\bb_{\Vb}, \cc_{\Ub }}
     {\gamma  (A, \bb ,\cc) 
     \leq     \frac{\epsilon }{\left(1 + \norm{\xxs}^{2} + \norm{\yys}^{2} \right)
                    \left(\norm{A} + 3 \right)}
\Big| \text{$A_{\Vb ,:} \xxs \leq \bb_{\Vb}$
                 and
               $\yys  A_{:,\Ub } \geq \cc_{\Ub}$}}.
\]
By Proposition~\ref{pro:max}, it suffices to
  prove that for all
  $U$, $V$,   $A_{\VUb}$, $\bb_{\Vb}$, $\cc_{\Ub}$,
  $\xxs$ and $\yys$,
  for which $  A_{\Vb ,U} \xxs_{U} \leq \bb_{\Vb }$ and
  $\yys_{V} A_{V, \Ub} \geq \cc_{\Ub}$,
\begin{equation}\label{eqn:probGamma1}
  \prob{A_{V,U}}
       {\gamma (A, \bb ,\cc ) \leq  \epsilon 
 } \leq 
\frac{
  \epsilon n e \left(1 + \norm{\xxs}^{2} + \norm{\yys}^{2} \right)
  \left(\norm{A_{V,U}} + 3 \right)
}{
  \sigma^{2}
},
\end{equation}
where we
  note that having fixed 
  $A_{\VUb}$, $\bb_{\Vb}$, $\cc_{\Ub}$,
  $\xxs$ and $\yys$,
  the induced distribution on $A_{V,U}$
  is
\begin{equation*}
   \mai{A_{V,U}} \mbi{A_{V,U} \xxs_{U}} \mci{\yys_{V} A_{V,U}}
   \det{A_{V,U}}^{2}.
\end{equation*}

To prove \eqref{eqn:probGamma1}, we show that
  for all $k \in U$ and all $A_{V,U-k}$,
\begin{equation}\label{eqn:probGamma2}
\prob{A_{V,k}}
     {\dist{A_{V,k}}{\Span{A_{V , U \setminus k}}} \leq 
      \epsilon }
\leq  
\frac{
  \epsilon e \left(1 + \norm{\xxs}^{2} + \norm{\yys}^{2} \right)
  \left(\norm{A_{V,U}} + 3 \right)
}{\sigma^{2}
  }
\end{equation}
and apply a union bound over $k \in U$.
Having fixed $A_{V,U-k}$, we may express
  $A_{V,k}$ as $\aa + t \qq $ where
  $\aa  \in \Span{A_{V,U-k}}$,
  $\qq$ is a unit vector orthogonal to 
  $\Span{A_{V,U-k}}$ and $t \in \Reals{}$.
With this representation, we have
  $\abs{t} = \dist{A_{V,k}}{\Span{A_{V , U \setminus k}}}$
  and $\det{A_{V,U}} = c t$,
  where $c$ is some constant
  depending only on $A_{V,U \setminus k}$.

By the symmetry of $\qq$ with $-\qq$, we can
  prove \eqref{eqn:probGamma2} by bounding
  the probability that $t$ is less than $\epsilon$
  given that $t$ is at least $0$.
Thus, we prove \eqref{eqn:probGamma2} by 
  observing $\norm{A_{V,U}} \geq \norm{A_{V,U-k},\aa }$
  and showing
\begin{equation}\label{eqn:probGamma3}
\max_{\aa  \in \Span{A_{V,U - k}}}
\prob{t}{t \leq  \epsilon \big| t \geq 0}
\leq  
\frac{
  \epsilon e \left(1 + \norm{\xxs}^{2} + \norm{\yys}^{2} \right)
  \left(\norm{A_{V,U-k}, \aa} + 3 \right)
}{\sigma^{2}
  }
\end{equation}
where the induced distribution on $t$ is proportional to
\begin{equation}\label{eqn:probGamma4}
  \rho (t) \defeq 
   \mu_{A_{V,k}} (\aa + t \qq ) 
  \mbi{A_{V,U-k} \xxs_{U-k} + (\aa + t \qq ) \xs_{j}}
  \mu_{c_{k}}\left(\yys_{V} (\aa + t \qq ) \right)
  t^{2}.
\end{equation}
We now set 
\[
\alpha = 
\frac{\sigma^{2}
}{
  3 \left(1 + \norm{\xxs}^{2} + \norm{\yys}^{2} \right)
  \left(\norm{A_{V,U-k}, \aa} + 3 \right)
}
\]
and prove
\begin{equation}\label{eqn:probGamma5}
  \text{$ 0 \leq t \leq t' \leq \alpha $ implies }
 \frac{\rho (t')}{\rho (t)} \geq 1/e,
\end{equation}
from which \eqref{eqn:probGamma3} follows by Lemma~\ref{lem:ratio}.

To prove \eqref{eqn:probGamma5}, we observe
\begin{itemize}
\item [1.]
As $\dist{\aa  + t' \qq}{\aa + t \qq} \leq t' - t \leq \alpha \leq 1$,
 we may apply Lemma~\ref{lem:GaussianSmooth} to show
\[
\frac{
  \mu_{A_{V,k}} (\aa + t' \qq)
}{
  \mu_{A_{V,k}} (\aa + t \qq)
}
\geq 
e^{\frac{-\alpha \left(\norm{\aa + t\qq} + 2 \right)}
        {\sigma^{2}}}
\geq 
e^{\frac{-\alpha \left(\norm{\aa} + 3 \right)}
        {\sigma^{2}}}
\geq 
e^{-1/3}.
\]

\item [2.]
As 
\[
\norm{A_{V,U-k} \xxs_{U - k}
      + \aa \xs_{k}
      + t \qq \xs_{k}}
\leq 
  \left(\norm{A_{V,U-k}, \aa} + \alpha \right) \norm{\xxs_{U}},
\]
and
\[
\dist{A_{V,U-k} \xxs_{U - k}
      + \aa \xs_{k}
      + t \qq \xs_{k}
}{
      A_{V,U-k} \xxs_{U - k}
      + \aa \xs_{k}
      + t' \qq \xs_{k}
}
= \left(t' - t \right) \xs_{k}
\leq \alpha \xs_{k},
\]
Lemma~\ref{lem:GaussianSmooth} implies
\[
\frac{
  \mbi{A_{V,U-k} \xxs_{U-k} + (\aa + t' \qq ) \xs_{i}}
}{
  \mbi{A_{V,U-k} \xxs_{U-k} + (\aa + t \qq ) \xs_{i}}
}
\geq 
e^{\frac{-\alpha \xs_{j} 
          \left( \left(\norm{A_{V,U-k}, \aa} + \alpha \right)
                  \norm{\xxs_{U}} + 2 \right)}
        {\sigma^{2}}}
\geq 
e^{-1/3}.
\]

\item [3.]
As $\abs{\yys (\aa + t \qq )} 
    \leq \norm{\yys}\left(\norm{\aa} + t \norm{\qq} \right)
    = \norm{\yys} (\norm{\aa} + \alpha )$,
and
\[
\dist{\yys (\aa + t \qq )}{\yys (\aa + t' \qq )}
= (t' - t) \abs{\yys \qq} 
\leq \alpha  \norm{\yys },
\]
Lemma~\ref{lem:GaussianSmooth} implies
\[
\frac{
  \mu_{c_{k}}\left(\yys_{V} (\aa + t' \qq ) \right)
}{
  \mu_{c_{k}}\left(\yys_{V} (\aa + t \qq ) \right)
}
\geq
e^{\frac{
- \alpha \norm{\yys}
  \left(\norm{\yys} (\norm{\aa} + \alpha ) + 2 \right)
}{
   \sigma^{2}
}}
\geq
e^{-1/3}.
\]
\end{itemize}
\end{proof}

\begin{proof}[Proof of Lemma~\ref{lem:probAlpha}
(Probability of small $\alpha $)]
By Lemma~\ref{lem:cov}, it suffices to bound
\[
\max_{U,V}
\prob{A, \xxs ,\yys ,\bb_{\Vb}, \cc_{\Ub }}
     {\alpha_{P}  (A, \bb ,\cc)
  \leq           \frac{\epsilon}{\left(\norm{A}+2 \right)^{2} 
                      \left(\norm{\xxs_{U}}+ 1 \right)}
\Big| \text{$A_{\Vb ,:} \xxs \leq \bb_{\Vb}$
                 and
               $\yys  A_{:,\Ub } \geq \cc_{\Ub}$}}.
\]
By Proposition~\ref{pro:max}, it suffices to
  prove that for all
  $U$, $V$, $A$, $\yys_{V}$ and $\cc_{\Ub }$,
  such that
  $\yys_{V} A_{V, \Ub} \geq \cc_{\Ub}$,
\begin{equation}\label{eqn:probAlpha0}
  \prob{\xxs_{U}, \bb_{\Vb }}
    {\alpha (A, \bb ,\cc) \leq 
            \frac{\epsilon}{\left(\norm{A}+2 \right)^{2} 
                      \left(\norm{\xxs_{U}}+ 1 \right)}
   \Big|  A_{\Vb ,U} \xxs_{U} \leq \bb_{\Vb }
     }
\leq 
  \frac{8 \epsilon n (m+1)}
       {\sigma^{2}},
\end{equation}
where we note that, fixing
  $U$, $V$, $A$, $\yys$ and $\cc_{\Ub}$ and conditioning
  upon $ A_{\Vb ,U} \xxs_{U} \leq \bb_{\Vb }$, the induced
  density on $\xxs_{U}$ is proportional to
\[
  \mbi{A_{V,U} \xxs_{U}}
  \prod_{j\not \in V}
  \prob{b_{j}}{b_{j} > A_{j,U}\xxs_{U} } .
\]
To prove \eqref{eqn:probAlpha0} , we show
\begin{align}
\forall_{i \in U} \forall_{\xxs_{U - i}}
  \prob{\xs_{i}}
    {\xs_{i} \leq   \epsilon
    \big| A_{\Vb ,U} \xxs_{U} \leq \bb_{\Vb } }
\leq 
  \frac{8 \epsilon 
        (m+1) \left(\norm{A}+2 \right)^{2} \left(\norm{\xxs_{U-i}}+ 1 \right)
        }
       {\sigma^{2}}, \label{eqn:probAlpha}
\end{align}
which implies
\[
\max_{i \in U} \max_{\xxs_{U - i}}
  \prob{\xs_{i}}
    {\xs_{i} \leq   
      \frac{\epsilon}{\left(\norm{A}+2 \right)^{2} 
                      \left(\norm{\xxs_{U-i}}+ 1 \right)}
    \Big|  A_{\Vb ,U} \xxs_{U} \leq \bb_{\Vb }}
\leq 
  \frac{8 \epsilon 
        (m+1)}
       {\sigma^{2}}.
\]
We then observe 
 $\frac{\epsilon}{\left(\norm{A}+2 \right)^{2} 
                      \left(\norm{\xxs_{U-i}}+ 1 \right)}
\leq 
\frac{\epsilon}{\left(\norm{A}+2 \right)^{2} 
                      \left(\norm{\xxs_{U}}+ 1 \right)}$
and union bound over $i \in U$.
To prove \eqref{eqn:probAlpha}, we first note that
  having fixed $i \in U$ and $\xxs_{U - i}$, the
  induced density on $\xs_{i}$ is proportional to
\[
\rho (\xs_{i}) \defeq 
  \mbi{A_{V,U-i} \xxs_{U - i} + A_{V,i} \xs_{i}}
  \prod_{j \not \in V}
  \prob{b_{j}}{b_{j} > A_{j,U-i} \xxs_{U - i} + A_{j,i} \xs_{i}}
\]
We now set 
\[
  \alpha 
 =
  \frac{\sigma^{2}}
       {4 (m+1)
        \left(\norm{A}+2 \right)^{2} 
        \left(\norm{\xxs_{U-i}}+ 1 \right)
        } \leq 1,
\]
and prove that
\begin{equation}\label{eqn:probAlpha2}
  \text{$ 0 \leq x_{i} \leq x_{i}' \leq \alpha $ implies }
 \frac{\rho (x_{i}')}{\rho (x_{i})} \geq 1/2,
\end{equation}
from which \eqref{eqn:probAlpha} follows by Lemma~\ref{lem:ratio}.

To prove \eqref{eqn:probAlpha2},
  we note that for $ 0 \leq x_{i} \leq x_{i}' \leq \alpha $,
\[
\dist{A_{V,U-i} \xxs_{U-i} + A_{V,i} x_{i}}
        {A_{V,U-i} \xxs_{U-i} + A_{V,i} x_{i}'}
 = 
 \norm{A_{V,i}} (x_{i}' - x_{i})
\leq 
 \norm{A} \alpha 
\leq 
1,
\]
and
\[
\norm{A_{V,U-i} \xxs_{U-i} + A_{V,i} x_{i}}
\leq 
\norm{A_{V,U-i}}
 \norm{\xxs_{U-i}}
+
\norm{A_{V,i}} x_{i}
\leq 
\norm{A} \left(\norm{\xxs_{U-i}} + 1 \right).
\]
So, by Lemma~\ref{lem:GaussianSmooth},
\[
\frac{\mbi{A_{V,U-i} \xxs_{U - i} + A_{V,i} x_{i}'}}
     {\mbi{A_{V,U-i} \xxs_{U - i} + A_{V,i} x_{i}}}
\geq 
e^{-\frac{\left(\norm{A} \left(\norm{\xxs_{U-i}} + 1 \right)+2 \right)
          \norm{A} \alpha 
         }
         {\sigma^{2}}
  }
\geq 
  e^{-\frac{1}{4 (m+1)}}
\geq 
  1 - \frac{1}{4 (m+1)},
\]
by our choice of $\alpha$.

We can also apply Lemma~\ref{lem:GaussianTails} to show
  that for each $j \not \in V$,
\begin{align*}
\frac{
  \prob{b_{j}}{b_{j} > A_{j,U-i} \xxs_{U - i} + A_{j,i} x'_{i}}
}{
  \prob{b_{j}}{b_{j} > A_{j,U-i} \xxs_{U - i} + A_{j,i} x_{i}}
}
& \geq 
1 - 
\frac{\left(\norm{A} \left(\norm{\xxs_{U-i}} + 1 \right) \right)
          \norm{A} \alpha 
         }
         {\sigma^{2}}
e^{\frac{\left(\norm{A} \left(\norm{\xxs_{U-i}} + 1 \right)+3 \right)
          \norm{A} \alpha 
         }
         {\sigma^{2}}
  }\\
& \geq 
1 - 
\frac{2 \left(\norm{A} \left(\norm{\xxs_{U-i}} + 1 \right) \right)
          \norm{A} \alpha 
         }
         {\sigma^{2}}\\
& \geq 
1 - 
\frac{1 }
         {2 (m+1)},
\end{align*}
by our choice of $\alpha$.
Thus, we may conclude
\[
 \frac{\rho (x'_{i})}
      {\rho (x_{i})}
 \geq 
\left(1 - \frac{1}{4 (m+1)} \right)
\left(1 - \frac{1 }{2 (m+1)} \right)^{m}
\geq 
1 - \frac{m+1}{2 (m+1)} = 1/2.
\]
\end{proof}

\section{Connection to Smoothed Analysis of Simplex Method}\label{sec:connection}

The analysis of the simplex method
  in~\cite{SpielmanTengSimplex} is broken into two parts:
  a combinatorial bound on the smoothed size of a two-dimensional shadow
  of a linear program, and
  an analysis of a two-phase algorithm that uses
  this combinatorial bound as a black-box.
The analysis of termination in this paper is closely related
  to the smoothed analysis of the shadow size.
The intuition behind this analysis is that if the
  angle at a corner of the polytope of feasible points
  is bounded away from being flat, then the simplex method
  should make significant progress as it traverses
  this corner.
The measure of angle used in~\cite{SpielmanTengSimplex}
  is approximately $\gamma (A, \bb ,\cc )$, at least
  for the corner optimizing the linear program.
The size of the shadow, which upper bounds the number
  of steps taken by the simplex method, is then bounded
  by varying $\cc$ over the plane onto which the shadow
  is projected.

The main technical lemma of the shadow-size analysis
  in~\cite{SpielmanTengSimplex}, Lemma~4.0.11 (Angle bound),
  essentially says that for every $\bb$, $\cc $ 
  and $\orig{A}$, the probability that a Gaussian
  perturbation $A$ of $\orig{A}$ has 
  $\alpha_{D} (A, \bb ,\cc) \gamma (A, \bb ,\cc) < \epsilon $
  is linear in $\epsilon$, with a coefficient polynomial
  in $n$, $m$ and $\sigma$.
The most significant difference between this statement and
  the analysis in Lemma~\ref{lem:probAlpha} and~\ref{lem:probGamma}
  is that in~\cite{SpielmanTengSimplex}, $\bb$ and $\cc$
  are not perturbed.
This restriction seems necessary to apply the combinatorial
  bound in a black-box fashion in the analysis of
  the two-phase simplex algorithm.
Also note that the simplex method analysis is for
  linear programs without the constraint $\xx \geq 0$.

Otherwise, the arguments in this paper have a flavor very
  similar to those of~\cite{SpielmanTengSimplex}, which
  mainly use the four techniques outlined in
  Section~\ref{sec:probabilistic} of this paper;
 although,
  that paper uses more elaborate changes of variables.
One probabilistic technique used in~\cite{SpielmanTengSimplex}
  that is absent in this paper is the 
  Combination Lemma~\cite[Lemma~2.3.5]{SpielmanTengSimplex}
  which allows one to obtain tight bounds on the probability that
  a product of parameters is small from bounds on the probabilities
  that the individual parameters are small.
The conditions of this lemma dictate the structure of the
  proofs in~\cite{SpielmanTengSimplex}
  as without it one could not obtain a bound on the probability
  of angle less than $\epsilon$ that is linear in $\epsilon$.
Moreover, without a bound that is linear in $\epsilon$, one could
  not prove that the shadow has expected polynomial size.
In contrast, in Lemma~\ref{lem:reduce} of this paper the dependency is
  on $\epsilon^{1/3}$.
It is possible that one could reduce this dependency using the
  combination lemma, but it is not essential for the results in
  this paper.

It is our hope that this paper will serve as a gentile introduction
  to the techniques used in the smoothed
  analysis of the simplex method.
  
\bibliographystyle{alpha}
\bibliography{ipm}

\end{document}